\documentclass[%
reprint,
superscriptaddress,
 amsmath,
 amssymb,
 aip,
 pop,
floatfix,
]{revtex4-2}

\usepackage{graphicx}
\usepackage{dcolumn}
\usepackage{bm}
\usepackage{hyperref}
\usepackage{gensymb}
\usepackage{xcolor}
\usepackage{algpseudocode}
\usepackage{algorithm}
\usepackage{lipsum}  
\usepackage{textgreek}

\hypersetup{
    colorlinks=true,
    linkcolor=blue,
    citecolor=blue, 
    urlcolor=black
    }
    
\begin{document}

\title{Evolution of laser-driven magnetic fields from proton tomography}

\author{J. Griff-McMahon}
\email{jgriffmc@pppl.gov}
\affiliation{Department of Astrophysical Sciences, Princeton University, Princeton, New Jersey 08544, USA\looseness=-1}
\affiliation{Princeton Plasma Physics Laboratory, Princeton, New Jersey 08540, USA\looseness=-1}

\author{V. Valenzuela-Villaseca}
\affiliation{Department of Astrophysical Sciences, Princeton University, Princeton, New Jersey 08544, USA\looseness=-1}
\affiliation{Lawrence Livermore National Laboratory, Livermore, California 94550, USA\looseness=-1}
\affiliation{Plasma Science and Fusion Center, Massachusetts Institute of Technology, Cambridge, Massachusetts 02139, USA\looseness=-1}

\author{C. A. Walsh}
\affiliation{Lawrence Livermore National Laboratory, Livermore, California 94550, USA\looseness=-1}

\author{S. Malko}
\affiliation{Princeton Plasma Physics Laboratory, Princeton, New Jersey 08540, USA\looseness=-1}

\author{B. McCluskey}
\affiliation{Department of Astrophysical Sciences, Princeton University, Princeton, New Jersey 08544, USA\looseness=-1}
\affiliation{Princeton Plasma Physics Laboratory, Princeton, New Jersey 08540, USA\looseness=-1}

\author{K. Lezhnin}
\affiliation{Princeton Plasma Physics Laboratory, Princeton, New Jersey 08540, USA\looseness=-1}

\author{H. Landsberger}
\affiliation{Department of Astrophysical Sciences, Princeton University, Princeton, New Jersey 08544, USA\looseness=-1}
\affiliation{Princeton Plasma Physics Laboratory, Princeton, New Jersey 08540, USA\looseness=-1}

\author{L. Berzak Hopkins}
\affiliation{Princeton Plasma Physics Laboratory, Princeton, New Jersey 08540, USA\looseness=-1}

\author{G. Fiksel}
\affiliation{Center for Ultrafast Optical Science, University of Michigan, Ann Arbor, Michigan 48109, USA\looseness=-1}

\author{M. J. Rosenberg}
\affiliation{Laboratory for Laser Energetics, University of Rochester, Rochester, New York 14623, USA\looseness=-1}

\author{D. B. Schaeffer}
\affiliation{Department of Physics and Astronomy, University of California Los Angeles, Los Angeles, California 90095, USA\looseness=-1}

\author{W. Fox}
\affiliation{Department of Physics, University of Maryland, College Park, Maryland 20742, USA\looseness=-1}
\affiliation{Princeton Plasma Physics Laboratory, Princeton, New Jersey 08540, USA\looseness=-1}

\begin{abstract}
Self-generated magnetic fields are commonly produced in high-power laser–plasma interactions. These fields can inhibit plasma heat-flow which makes them important in inertial fusion and controlled laboratory astrophysics experiments. In this work, we characterize the time evolution of self-generated magnetic fields using multi-view proton tomography at two timings. Tomographic reconstructions of the magnetic field show a clear transition from fields located close to the target at early time to more extended coronal fields at later time. The tomographic inversion and mesh radiography also enable a direct measurement of the magnetic-flux evolution. Comparisons with extended-MHD simulations show only moderate agreement in field structure, but good agreement in magnetic flux. This suggests that the field generation model is largely correct under these conditions, while the magnetic transport model requires additional development to reproduce the observed field structure.

\end{abstract}

\maketitle

\section{Introduction}

Self-generated magnetic fields ubiquitously form in high-power laser-solid interactions through the Biermann-battery effect \cite{biermann_uber_1950,stamper_spontaneous_1971}.
In this effect, a pressure-gradient driven EMF (electromotive force) with non-vanishing curl produces a ``battery" that drives loop currents and generates magnetic fields in the plasma. The magnetohydrodynamic description of Biermann-battery magnetic field generation is
\begin{equation}
    \frac{\partial B_{\rm{Biermann}}}{\partial t}=\nabla \times \left(\frac{\nabla P_e}{e n_e}\right)
\end{equation}
where $P_e$ is the electron pressure, $e$ is the fundamental charge, and $n_e$ is the electron density. In laser-produced plasmas, magnetic fields then evolve through advection by electron fluid motion and thermoelectric forces as well as dissipation by ohmic heating \cite{braginskii_transport_1965,walsh_extended-magnetohydrodynamics_2020}.

A key consequence of magnetic fields is the modification of plasma transport through insulation of electron heat flow perpendicular to the field \cite{froula_quenching_2007} and deflection of electron heat flow along isotherms through the Righi-Leduc effect \cite{sadler_symmetric_2021}.
Several recent simulation studies have highlighted the importance of self-generated magnetic fields in laser-driven hohlraums, used for indirect-drive inertial confinement fusion (ICF) \cite{abu-shawareb_achievement_2024}. Even modest magnetization in the blowoff plasma or around the laser-entrance hole of the hohlraum can substantially reshape the temperature profile and modify laser-plasma coupling \cite{farmer_simulation_2017,leal_effect_2025}, implosion symmetry \cite{moody_increased_2022,bose_effect_2022} and laser-plasma instabilities \cite{shi_particle--cell_2025}.
In parallel, laser-plasmas are well-suited to perform a variety of laboratory astrophysics experiments including studies of magnetic reconnection \cite{nilson_bidirectional_2008,rosenberg_slowing_2015,valenzuela-villaseca_x-ray_2024}, various magnetic instabilities \cite{gao_magnetic_2012,manuel_first_2012,fox_filamentation_2013}, astrophysical jets \cite{fu_creation_2015,gao_mega-gauss_2019}, and dynamo amplification of magnetic fields \cite{tzeferacos_laboratory_2018,bott_time-resolved_2021}. A full understanding of magnetic field generation and transport in laser-plasma interactions is important for predictive models of laser fusion and controlled laboratory astrophysics experiments.

Although the generation and transport of magnetic fields in laser-plasmas has been studied for over half a century \cite{stamper_spontaneous_1971}, there are still key questions that remain unresolved. For example, there is no clear consensus on where the fields actually reside in a laser-solid interaction. For the past two decades, magnetohydrodynamic (MHD) and extended MHD simulations have predicted a wide range of morphologies for the Biermann battery-driven magnetic fields, ranging from expanding shells frozen into the edge of the coronal plasma \cite{li_measuring_2006}, to thin ``pancake" structures compressed down near the solid density surface due to the Nernst effect \cite{lancia_topology_2014,gao_precision_2015,campbell_measuring_2022}. The Nernst effect advects magnetic fields down electron temperature gradients with the electron heat flux, and is often thought to anchor and compress the fields close to the cold target \cite{nishiguchi_convective_1984}. However, recent experimental measurements\cite{griff-mcmahon_structure_2025} and results presented in this paper observe fields that extend into the corona, suggesting that Nernst anchoring may be more limited than previously thought. These differences are significant because they impact whether the field magnetizes the coronal plasma, and has any impact at all on processes such as the plasma heat transport \cite{lancia_topology_2014}.

Another key question is how much total magnetic flux ($\Psi = \int \mathbf{B} \cdot d\mathbf{A}$) is generated in laser-solid interactions. Recent kinetic simulations have predicted that Biermann-battery field generation is suppressed by nonlocal effects that occur when the electron mean free path approaches the electron temperature gradient length scale ($\lambda_{ei}/L_T \gtrsim0.01$) \cite{ridgers_inadequacy_2020,sherlock_suppression_2020,davies_nonlocal_2023}. Extended MHD simulations with Biermann suppression models predict a factor of 2--3 reduction in magnetic flux in laser-solid interactions \cite{campbell_measuring_2022,griff-mcmahon_measurements_2024}. However, existing measurements of the magnetic flux span a broad range; 
some measurements have provided evidence for suppression of field generation\cite{li_observation_2007,campbell_measuring_2022} whereas others \cite{griff-mcmahon_measurements_2024,griff-mcmahon_structure_2025} and the present results show best agreement with simulations that do not include suppression. Accurate measurements of magnetic field structure and magnetic flux can constrain models of magnetic field generation and transport. 

Characterization of self-generated magnetic fields has typically relied on proton radiography \cite{schaeffer_proton_2023} from a single line of sight, providing path-integrated measurements of magnetic fields through the plasma \cite{li_observation_2007,petrasso_lorentz_2009,willingale_fast_2010,gao_precision_2015,campbell_magnetic_2020,campbell_measuring_2022,griff-mcmahon_measurements_2024}. However, the path-integrated nature of the diagnostic has prevented a full understanding of where the field is located along the proton probing axis. To address this limitation, we developed a multi-view proton tomography method to infer the 3D field structure and magnitude of self-generated magnetic fields in a laser-solid interaction, rather than the typical path-integrated quantities \cite{griff-mcmahon_structure_2025}.

This paper has two primary aims. First, we present the tomography inversion methodology in detail, including careful discussion of the requirements and the associated statistical and robustness tests we have developed. Second, we then extend the results of Ref. \cite{griff-mcmahon_structure_2025} by applying the tomography method to measure the evolution of self-generated magnetic field structure at two probe times. At each timing, we image the system with mesh proton radiography with either two or four view angles to enable a tomographic inversion. At early time ($t=0.7~$ns), the fields are located near the target and become extended off the target at late time ($t=1.4~$ns). The extended fields are strong enough to magnetize the coronal plasma and would be sufficient to suppress cross-field heat transport. This is directly relevant to ICF hohlraums, where self-generated magnetic fields in the wall blowoff and near the laser-entrance hole can modify electron heat flow, laser propagation, and drive symmetry. The tomographic reconstructions and advanced mesh radiography methods also enable a direct measurement of the magnetic flux evolution that isolates field generation. We compare the magnetic structure and flux to extended MHD simulations that include a recent Biermann-battery suppression model that accounts for re-localization by the magnetic fields. While the magnetic flux is well matched, the field structure shows only moderate agreement between experiment and simulation. This suggests that the net Biermann-battery generation is captured at the correct level, whereas the magnetic transport models remain incomplete. The remainder of this paper describes the experimental platform and inversion methodology in detail.

\section{Methods}

\begin{figure}
	\includegraphics[width=\linewidth]{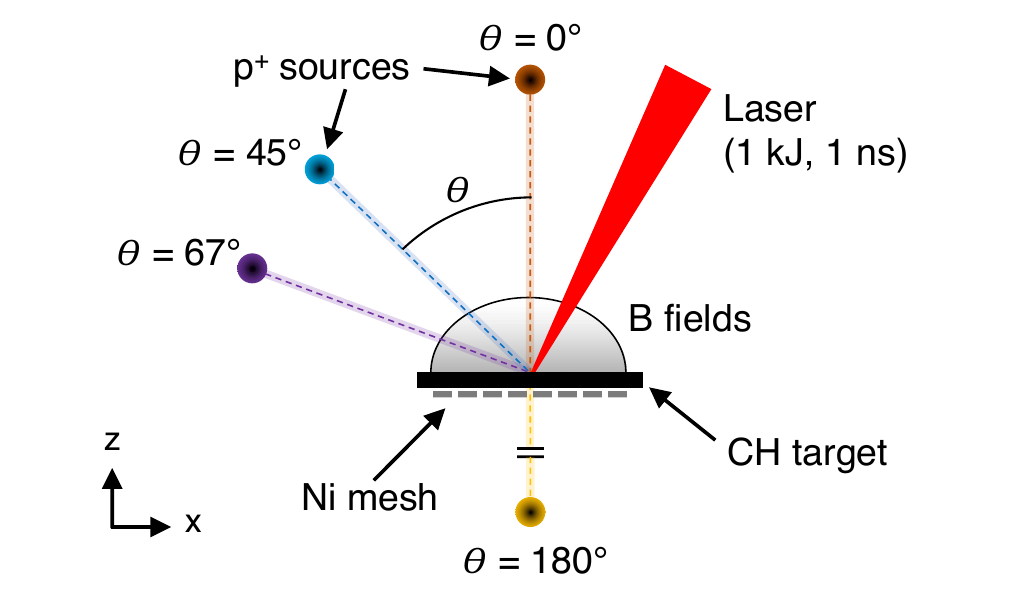}
	\caption{Experimental setup with different proton backlighter source positions (colored circles). The backlighter angle $\theta$ is defined relative to the target normal. The $t=0.7~$ns timing used 0\degree and 45\degree views, while the $t=1.4~$ns timing used 0\degree, 45\degree, 67 \degree and 180\degree views.}
	\label{fig:setup}
\end{figure}

\begin{figure*}
	\includegraphics[width=\linewidth]{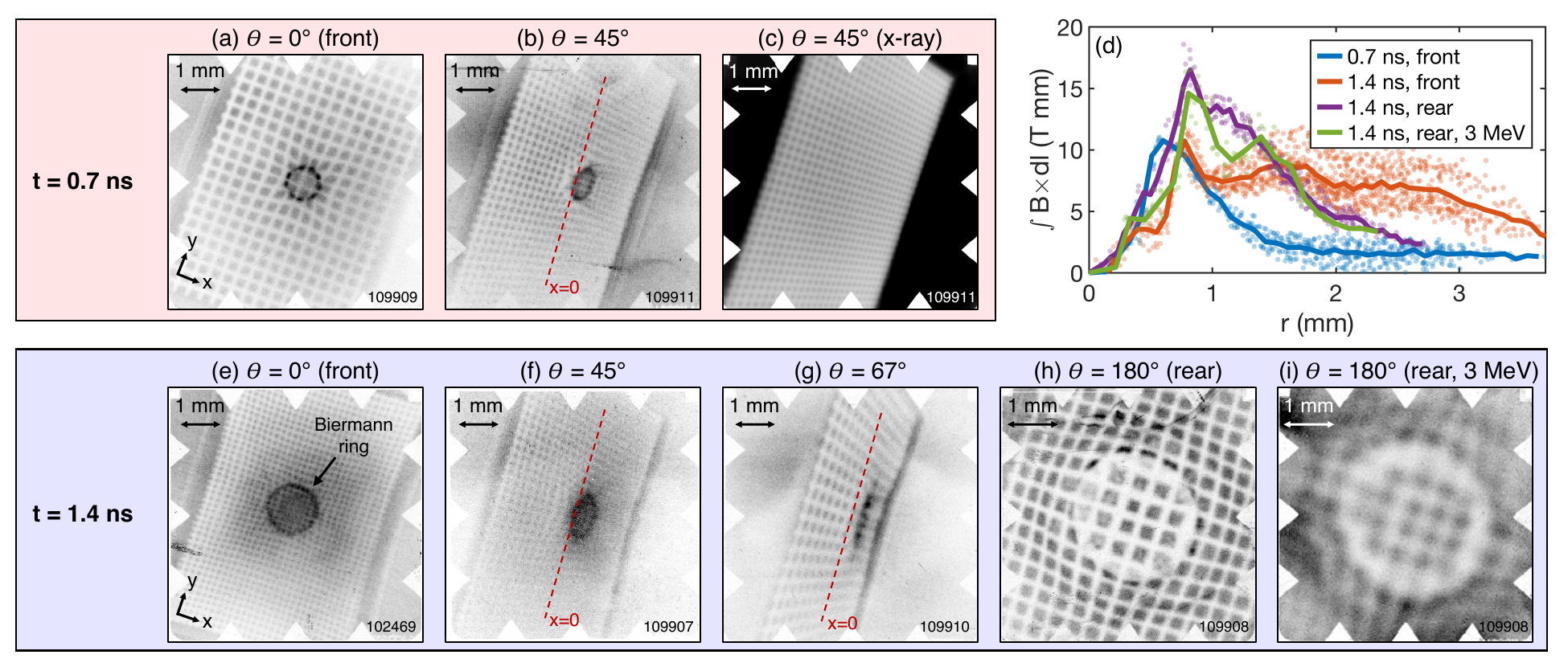}
	\caption{Experimental proton radiographs at $t=0.7~$ns (top row) and $t=1.4~$ns (bottom row) as the target is rotated about the $y$-axis in successive shots. All images show 15 MeV protons except (c) which is the x-ray reference image for 45$\degree$ and (i) which is a 3 MeV proton image. Darker regions received higher fluence. A clear shift is visible between the Biermann ring and the $x=0$ axis aligned with the laser spot in (b,f,g). (d) Path-integrated magnetic field profiles from the front and rear views at the two timings. Each dot is a beamlet deflection and the line is the azimuthally-averaged value.}
	\label{fig:data_comb}
\end{figure*}

\subsection{Experimental setup and data}

This experimental campaign was performed on the OMEGA laser at the Laboratory for Laser Energetics at the University of Rochester. To produce the magnetic fields, two co-timed drive beams irradiated a 25-$\mu$m thick CH foil. The foil has dimensions $5\times6.6~$mm. The beams were 351~nm wavelength and delivered a combined laser energy of 1 kJ with 1 ns duration in a square pulse shape. The $1/e$ intensity radius was 358 $\mu$m using an SG5 distributed phase plate \cite{lin_distributed_1995}, corresponding to peak intensity of $3\times10^{14}~$W/cm$^2$. A backlighter capsule filled with D$^3$He gas was irradiated with a separate set of 26 beams to provide a source of fusion-product protons with monoenergetic populations at 3 and 15 MeV energies from the D+$^3$He and D+D fusion reactions. Each beam to the backlighter was 500 J and 1 ns pulse duration. The capsule was positioned 10 mm away from the foil center. A CR-39 detector stack measured the proton fluence at a standoff of either 154 or 214 mm from the foil, resulting in 16.4$\times$ or $22.4\times$ image magnification, respectively. 
A nickel mesh was attached to the rear surface of the target. As the protons passed through the mesh holes, they formed discrete beamlets whose electromagnetic deflections could be tracked.
X-rays emitted from the backlighter were also measured on an Image Plate (IP) at the rear of the detector stack and used to provide a reference image of the mesh for each shot \cite{johnson_proton_2022,malko_design_2022}. The beamlet deflections $\vec d$ provide a direct measurement of the electromagnetic field strength integrated along each proton trajectory.
\begin{equation} \label{eq:deflection}
    \vec d = \frac{e L}{m_p v_p^2}\int dl\, G \left[\vec E + \vec v_p \times \vec B \right]_\perp
\end{equation}
Here, $e$ is the fundamental charge, $L$ is the distance from the target to the detector, $m_p$ is the proton mass, $v_p$ is the proton velocity, $G$ is a geometric correction factor that accounts for proton deflections that occur before reaching the mesh, discussed later in this section and in Appendix \ref{sec:geometric_corr}, and the $\perp$ symbol references perpendicular to the proton trajectory.

To perform proton tomography, we imaged the interaction from multiple view angles in successive shots as shown in Fig. \ref{fig:setup}. We probed at $t=0.7~$ns using two different backlighter positions at $0\degree$ and $45\degree$, where $t=0$ refers to the start of the laser drive. We also probed at a later timing of $t=1.4~$ns using $0\degree$, $45\degree$, $67\degree$, and $180\degree$ views. In practice, the backlighter source and detector were kept fixed and the target was rotated from shot-to-shot. We take advantage of the 60 available beams on OMEGA to select different drive beams for each view angle. This maintains a similarly small angle-of-incidence for all shots and produces comparable fields, discussed in more detail later in Sec. \ref{sec:inv_assumptions}.

Figure \ref{fig:data_comb} shows experimental proton radiographs at $t=0.7~$ns (top row) and $t=1.4$ ns (bottom row), as the target was rotated about the $y$-axis in successive shots. Here, the $y$-axis is aligned with the longer dimension of the foil (nearly vertical in the radiographs in Fig. \ref{fig:data_comb}) and passes through the center of the foil. Each proton image has a corresponding x-ray reference image, of which an example pair is shown in Figs. \ref{fig:data_comb}(b,c). The jagged fiducial borders on the x-ray and proton images are used to align the images and enable pointwise extraction of proton deflections through the mesh. All proton radiographs used 15 MeV protons, except for Fig. \ref{fig:data_comb}(i) which used 3 MeV protons. The combination of 3 and 15 MeV proton images from the same view distinguishes between electric and magnetic fields \cite{griff-mcmahon_measurements_2024,pearcy_revealing_2025}.

There are several aspects to note across the radiographs in Fig. \ref{fig:data_comb} from the viewing geometry and diagnostic configuration. The 0$\degree$ and 180$\degree$ views in Fig. \ref{fig:data_comb}(e,h) show opposite polarity of proton deflections because the sign of the Lorentz force reverses when the line-of-sight is flipped; 0$\degree$ focuses protons inwards while 180$\degree$ deflects protons outwards. The 180$\degree$ view also fielded an increased detector standoff (and thus larger magnification), reducing the diagnostic field of view. The 3 MeV image in Fig. \ref{fig:data_comb}(i) is blurred relative to the 15 MeV images due to increased proton scattering in the target and mesh, however the beamlets remain clearly visible. 

In addition, different nickel meshes were used to optimize contrast and spatial sampling for different shots. Figure \ref{fig:data_comb}(a,g-i) used a coarse mesh (306$~\mu$m cell size and 70$~\mu$m thickness) to provide high contrast at reduced nominal spatial resolution. However, in practice both the mesh holes and cross-bars were used to determine beamlet deflections for Fig. \ref{fig:data_comb}(a,h), which doubled the spatial resolution. Figure \ref{fig:data_comb}(b,c,f) used a medium-coarse mesh (175$~\mu$m cell size and 75$~\mu$m thickness) and Figure \ref{fig:data_comb}(e) used a fine mesh (150$~\mu$m cell size and 60$~\mu$m thickness). The thickness of the mesh was important for producing high-contrast radiographs. In all cases, the mesh was glued to the rear surface of the target. 

\subsection{Qualitative tomography} \label{sec:qual_tomo}

Before performing the full tomographic inversion, it is instructive to conduct simpler analyses at $t=1.4~$ns to build intuition and recover qualitative information about the field structure. For example, tomographic information is encoded in the ``Biermann ring" feature, labeled in Fig. \ref{fig:data_comb}(e), where protons are focused inwards onto each other. As the view angle becomes shallower [Fig. \ref{fig:data_comb}(e-g)], this feature shifts in the positive $x$-direction relative to the reference $x=0$ axis passing through the laser spot. This indicates that the magnetic fields are located at some height above the target surface. If instead the fields were confined to the target surface, the Biermann ring would remain stationary as the views changed. This immediately indicates that the tomographic views contain information about the 3-D field structure.

In addition, a comparison between the front and rear views and between different proton energies gives further information about the field structure and whether deflections originate from electric or magnetic fields. Figure \ref{fig:data_comb}(d) shows radial profiles of the path-integrated magnetic field for the front and rear views at both timings. These use Eq. \eqref{eq:deflection} assuming that the deflection is a result of magnetic fields only ($ E_\perp=0$) and that there is no geometric correction ($G=1$). The radial coordinate is defined by where the x-rays from the backlighter intersect with the target plane. The 3 and 15 MeV magnetic profiles from the same shot [purple and green lines in Fig. \ref{fig:data_comb}(d)] are nearly identical. Since magnetic and electric fields have different proton energy scalings, any appreciable radial electric fields would produce a separation between these curves. Their agreement therefore indicates that radial electric fields are subdominant and that the deflections are primarily magnetic $(E_r \ll v_pB)$. This finding agrees with previous analyses \cite{petrasso_lorentz_2009,griff-mcmahon_measurements_2024}.

Comparison of the front and rear views [orange and purple lines in Fig. \ref{fig:data_comb}(d)] also provides information about the spatial location of the magnetic field. This is because the two views probe different regions along a given proton path and because the inferred deflection depends on the geometric mesh correction, $G=1-L_{\rm off}/L_{\rm mesh}$, used in Eq. \eqref{eq:deflection} for protons deflected before reaching the mesh. Here, $L_{\rm off}$ is the distance from the field to the mesh and $L_{\rm mesh}$ is the distance from the source to the mesh. As a result, protons and x rays passing through the same mesh hole can take slightly different initial trajectories, which changes the deflection angle inferred at the detector \cite{griff-mcmahon_structure_2025}. Again, we note that if the $B$ fields were fully pancaked onto the target surface, then the front and rear views would produce identical radial profiles. Additional details are provided in Appendix \ref{sec:geometric_corr}.

Returning to the data, the fact that 0$\degree$ (orange) and 180$\degree$ (purple) views observe different line integrated fields, for example with the peak field reduced by $\sim$35\% when viewed from the front, again indicates the different views are providing data about the 3-D field structure. Since we showed above that the radial electric fields are not significant, the difference is likely from the geometric factor. In the rear view, the protons pass first through the mesh and then the field region so there is no correction (Fig. \ref{fig:setup}). In contrast, the front view deflections occur upstream of the mesh so the geometric correction must be applied. The reduction in the peak path-integrated field in the front view suggests that significant magnetic fields lie above the target surface. The inversion methodology will now synthesize this and the remaining data to constrain the 3-D field structure.

\subsection{Inversion assumptions and framework} \label{sec:inv_assumptions}

\begin{figure}
	\includegraphics[width=\linewidth]{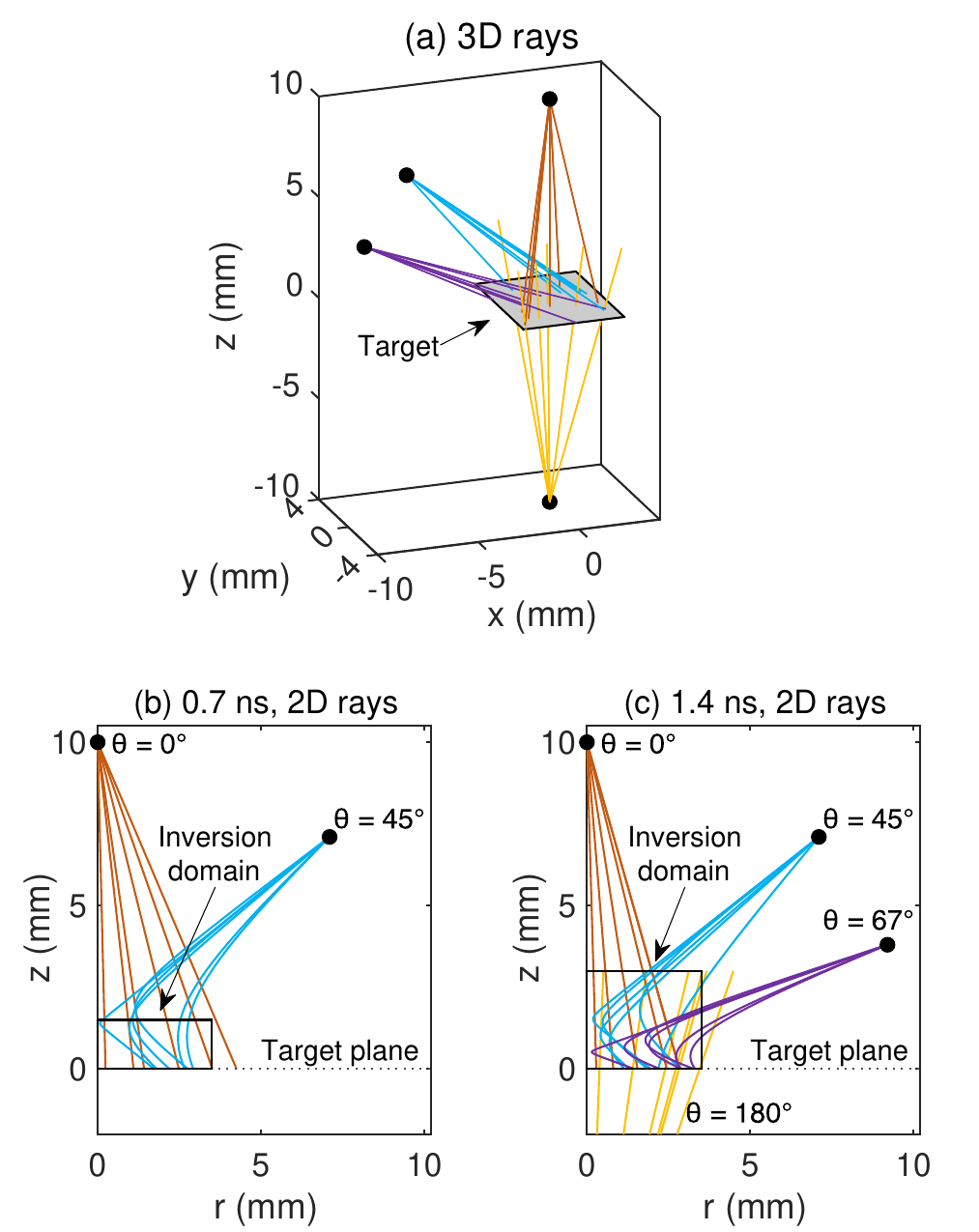}
	\caption{(a) Sample straight-line proton trajectories from the four proton sources (black dots) in 3D. (b,c) Trajectories projected onto cylindrical coordinates for $t=0.7~$ns and $t=1.4~$ns. The inversion domain is shown in by the black rectangle.}
	\label{fig:RZ_rays_comb}
\end{figure}

We now present the assumptions and framework used to perform a quantitative tomographic inversion to recover the toroidal magnetic field $B_\phi$. We do not include electric fields as they were found to be subdominant in the prior analysis. In fact, we also explored tomography including electric fields, but found they did not significantly change the magnetic structure or integrated quantities like magnetic flux. The inversion relies on three main assumptions: shot-to-shot reproducibility, axisymmetry, and small-angle deflections.

First, shot-to-shot reproducibility is required because the tomographic reconstruction combines multiple shots at different view angles. To ensure reproducibility, different laser beams were selected for each view angle to maintain a similar angle of incidence as the target was tilted in successive shots. Across the four views, the angles of incidence ranged from  21.3$\degree$ to 25.5$\degree$ from target normal, producing a single-beam ellipticity of 7 to 11\%. However, the use of two overlapped beams per laser spot from different azimuthal angles partially compensated for this effect and yielded a repeatable, circular beam profile. Similarly, the laser energy varied only slightly across the six shots and ranged from 965 to 1000 J on target, corresponding to less than 3.5\% relative change. Lastly, the target tilt was known to within 2$\degree$ and enabled good calibration of the proton source relative to the target and the laser beam spot. 

Second, we assume axisymmetry in the toroidal magnetic field $B_\phi = B_\phi (r,z)$. The small amount of non-axisymmetry is visible through the scatter in the radial profiles in Fig. \ref{fig:data_comb}(d). The azimuthally averaged standard deviation is less than 1.5$~$T$\,$mm for all shots, corresponding to around 20\% variation for the front view at $t=1.4~$ns and around 10\% variation for the other shots. These variations might stem from small amounts of ellipticity in the beam spot or slight target warping, but are sufficiently small to invoke axisymmetry. This assumption enables projection of the proton trajectories onto cylindrical coordinates and increases the density of the tomographic coverage. Figure \ref{fig:RZ_rays_comb}(a) shows representative straight-line trajectories from each proton source through the inversion domain. Each ray corresponds to a single mesh measurement. The corresponding projected trajectories in cylindrical coordinates are shown in Fig. \ref{fig:RZ_rays_comb}(b,c) for the two timings. We emphasize that the apparent curvature of the trajectories from the oblique views is not due to proton deflection, but is simply a projection of straight-line trajectories onto cylindrical coordinates. The overlap of multiple trajectories in Fig. \ref{fig:RZ_rays_comb}(b,c) provides information about the structure of the fields. Despite using only two views at $t=0.7~$ns, the cylindrical projection leads to coverage over the inversion domain, as shown in black boxes in Fig. \ref{fig:RZ_rays_comb}(b,c). 

The final assumption is that the proton deflections are small enough that protons effectively sample the fields along the straight-line x-ray trajectories. This establishes which inversion cells contribute to the deflection of a single proton. This approximation holds when the transverse deflection in the domain is smaller than the inversion cell size, which we take to be $150~\mu$m in these inversions. This sets a lower bound on the inversion cell size. Using a deflection angle of $\alpha=\frac{e}{m_p v_p} \int B\,dl$, the worst-case intra-domain transverse displacement is $\Delta=\alpha \ell$ for path length $\ell$ in the domain. For peak path-integrated field of $15~$T$\,$mm measured in \ref{fig:data_comb}(d), $\alpha \approx 1.5\degree$ for 15 MeV protons and this approximation is satisfied for $\ell<5.5~$mm. In addition, we validate the final inversion by forward-modeling proton trajectories that include intra-domain deflections. Future work may develop a fully self-consistent iterative solver that updates proton trajectories through the reconstructed fields, with applications to systems with stronger magnetic fields.

\subsection{Inversion methodology} \label{sec:tomographic_inversion}

This section describes the tomographic inversion used to infer the toroidal magnetic field from measured proton deflections. We use an algebraic reconstruction framework \cite{kak_principles_2001} that represents the magnetic deflection $\vec d$ as a sum of small deflections along the proton path and solves for $B_\phi(r,z)$ iteratively.

\begin{figure}
	\includegraphics[width=\linewidth]{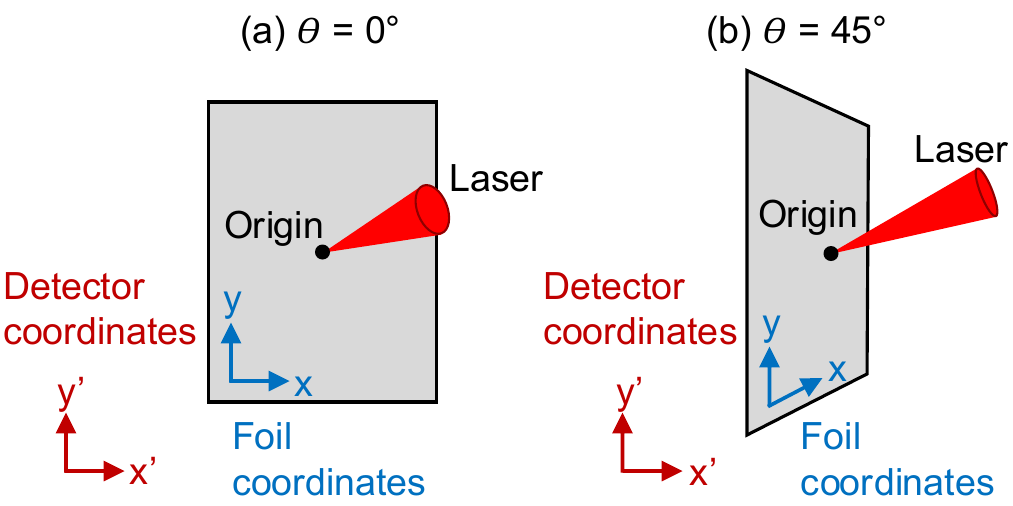}
	\caption{Detector and foil coordinate systems for tilt angles of (a) $\theta=0\degree$ and (b) $\theta=45\degree$.}
	\label{fig:coord_sys}
\end{figure}

First, we define a common target coordinate system $\vec x = (x,y,z)$ where the $xy$-plane lies on the foil and the $z$-axis is normal to the target, as shown in Fig. \ref{fig:coord_sys}. Each view also has a detector coordinate system $\vec x' = (x',y',z')$ whose $z'$-axis is aligned with the proton backlighter axis, and is related to the target frame by a rotation of $\theta$ about the $y$-axis. For each mesh measurement in the x-ray image, a straight-line ray is constructed from the backlighter source to the corresponding point on the target plane using the known target tilt and system geometry. We parameterize each ray's trajectory as $\vec x(l) = \vec x_0 + \hat v_p l$ where $\vec x_0$ is the proton intersection point on the target plane, $\hat v_p$ is the proton velocity direction, and $l$ is the path length.

The field is purely toroidal in the target coordinates so that $\vec B = B_\phi(r,z)\hat \phi$. The total deflection is represented as a sum of single-cell contributions:

\begin{equation}
    \vec d = \frac{e L}{m_p v_p^2} \sum_k G_k dl_k B_k\, [\vec v_p \times \hat \phi].
\end{equation}
Here, the index $k$ refers to the $k^{th}$ cell in the domain, $dl_k$ is the path length in that cell, and $G_k$ is the mesh correction based on the distance between the $k^{th}$ inversion cell and the proton intersection with the target plane. The crux of this technique is (1) constructing the proton trajectory through the inversion domain and (2) computing each cell's contribution to the measured deflection.

We decompose the detector deflection $\vec d$ into an orthogonal basis defined in detector coordinates. Natural choices are a Cartesian basis $(x',y')$ or a cylindrical basis $(r',\phi')$. Although both are valid, we use the cylindrical basis because front and rear views should deflect purely along $r'$ and because this choice extends naturally beyond the paraxial approximation of Eq.~\eqref{eq:deflection}, discussed later.

The total deflection is a linear combination of deflections from each cell $k$ the proton passes through. The deflection for the $i^{th}$ ray in the basis vector $\hat {e} = (\hat{r'},\hat{\phi'})$ direction is
\begin{align} 
\label{eq:deflection_basis} 
d_{i}^{(\hat e)} &= \sum_k \mathbf M_{ik} B_k,\\ 
\mathbf M_{ik} &= \frac{e L}{m_p v_p^2} G_k dl_k\left[ (\vec v_p \times \hat \phi) \cdot \hat {e} \right] (\sec{\theta'})^a.
\end{align}

Each cell in the inversion contributes a weight of $\mathbf M_{ik}$ to the total deflection, resulting in a sparse population of $\mathbf{M}$ because each proton traverses only a small subset of inversion cells. The deflections are projected onto the $\hat {e}$ direction and $\theta'$ is defined as the polar angle between the proton ray and the detector normal. The $\sec{\theta'}$ factor has exponent $a=(3,1)$ for $\hat e = (r',\phi')$, and is a non-paraxial correction that accounts for increased path-length in the fields and geometric projection on the detector. Bilinear interpolated weighting is used to distribute each ray's path length among the four neighboring cells to reflect the sub-cell position rather than a nearest-cell assignment. This increases the accuracy of the inversion, as tested with synthetic data. 

A linear system of equations is constructed from all the measured deflections 
$\mathbf{M}_{2m \times n}\vec B_{n \times 1}=\vec d_{2m \times 1}$ where $\mathbf{M}_{2m \times n}$ is the coefficient matrix containing information about the ray trajectory and geometric factors, $\vec B_{n \times 1}$ is a vector containing $n$ magnetic field cells in the inversion, and $\vec d_{2m \times 1}$ is a vector of $m$ measured deflections along two basis dimensions. Equivalently, each of the $m$ rays contributes two scalar equations, one for each deflection component. The system is then solved for the magnetic field through an iterative least squares solver using the simultaneous iterative reconstruction technique (SIRT) \cite{hansen_air_2018}. Although algebraic reconstruction techniques are typically applied to scalar tomography, the same linear framework applies to the vector deflection because each deflection component acts as a scalar constraint.

We also enforce a sign constraint ($B_\phi<0$) to suppress small patches of field reversal that arise from both noise and limited tomographic coverage in portions of the domain. Adding a sign constraint has only a minor impact on the final inversion and does not significantly affect integrated quantities such as path-integrated field profiles or the magnetic flux. It is also straightforward to extend this approach to include additional fields like the radial electric field and $z$-directed electric fields based on Eq. \eqref{eq:deflection_basis}. A validation of this technique with synthetic data is shown in Appendix \ref{sec:Tomo_Validation}.

\begin{figure}
	\includegraphics[width=\linewidth]{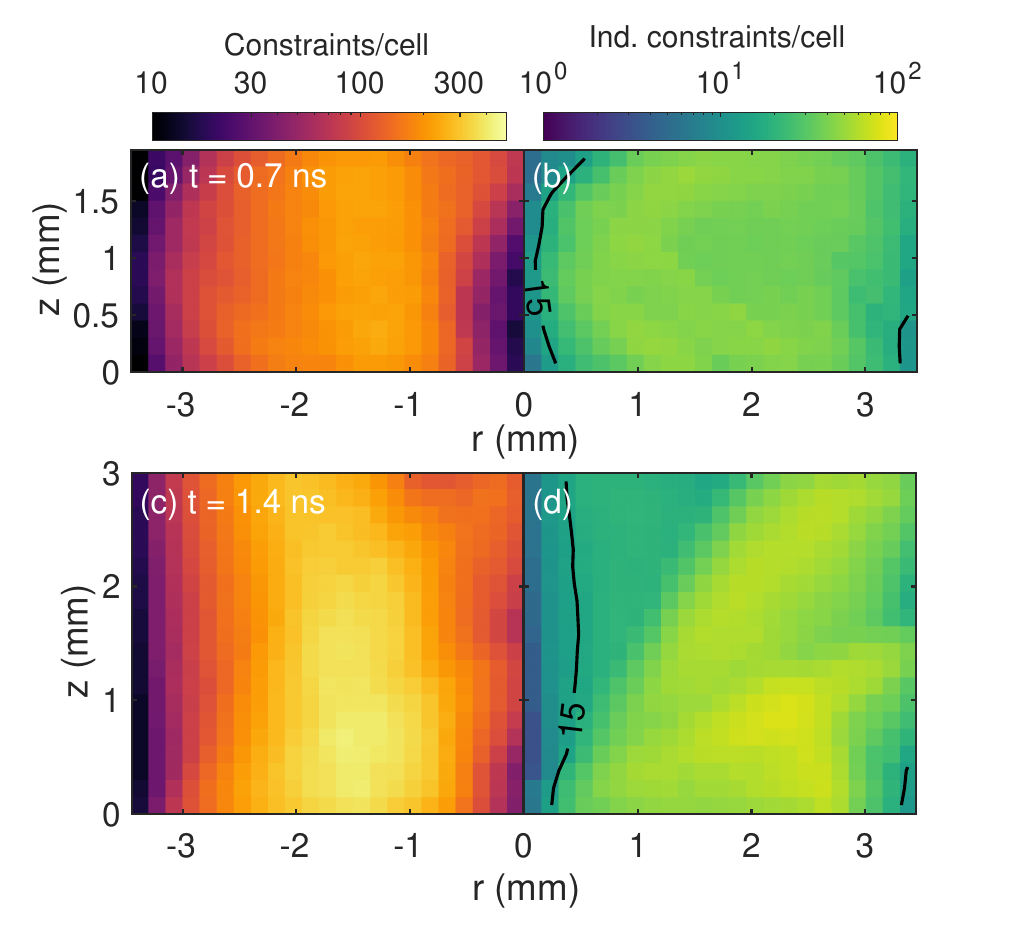}
	\caption{(a,c) Number of constraints in each cell. (b,d) Number of independent constraints in each cell. The black contour line indicates 15 constraints.}
	\label{fig:Tomo_coverage}
\end{figure}

The coverage and conditioning of the system are summarized in Fig.~(\ref{fig:Tomo_coverage}). The left column shows the number of scalar deflection measurements from protons that pass through an inversion cell, at both timings. For oblique views, each beamlet contributes two scalar deflection constraints for each cell that the proton intersects with. Beamlets in the front and rear views contribute only one radial measurement in each trajectory cell since azimuthal deflections are not modeled in this system. The central region for each timing has over 200 overlaps across a range of proton trajectories, which averages over the noise and stabilizes the fit. The inversion is highly over-constrained since there are many more proton deflection measurements than inversion cells, with 2266 scalar constraints and 230 cells at $t=0.7~$ns, and 4232 constraints and 460 cells at $t=1.4~$ns. There are approximately 10 constraints per unknown. This number is influenced by the inversion cell size of 150~$\mu$m and the mesh cell size of 150 to 306~$\mu$m. 

On its own, an overdetermined system does not guarantee a well-conditioned inversion. In fact, the raw number of constraints per cell overstates the amount of independent information because many measurements are geometrically degenerate. For example, in the front view (0$\degree$), beamlets that intersect the target with the same radius follow the same trajectory in $(r,z)$ space. They therefore produce the same linear constraint on $B_\phi$ and do not contribute new independent information. These redundant measurements primarily act to average over the noise, but do not further constrain the field structure. The oblique-angle views provide the non-degeneracy in ray geometry that is needed to localize $B_\phi$.

To quantify redundancy, Figs.~\ref{fig:Tomo_coverage}(b,d) show an ``effective'' number of independent constraints per cell. We first compute the singular value decomposition $\mathbf{M}=\mathbf{U}\mathbf{\Sigma}\mathbf{V}^T$ to determine the matrix rank $r$ using a relative threshold of $10^{-4}$ on the singular values. We then select $r$ measurements that are maximally independent using QR factorization with column pivoting \cite{golub_matrix_2013}. This selects the $r$ rows in $\mathbf{U}$ that are as close as possible to orthogonal. This produces a representative set that spans the same information as the full dataset but discards linearly dependent (duplicated) measurements. Finally, we map these selected measurements into the inversion domain by counting how many of the selected rays intersect each cell. This measures the number of independent constraints on that cell and shows that the central region is substantially better constrained than the periphery. In particular, the small-radius, large-$z$ region is sampled by only a few independent rays and therefore carries the largest uncertainty. 

\section{Tomography results}

This section first discusses the magnetic structure inferred in the laser-solid interaction experiment and its implications. Then, we compare the experimental fields to extended MHD simulations and find moderate agreement in field structure. Next, we discuss the magnetic flux, which isolates net field generation, and agrees well with simulations that include a re-localized model of Biermann-battery field suppression. Lastly, we perform several validation analyses to support the fidelity of the inversion.

\subsection{Magnetic structure}

\begin{figure}
	\includegraphics[width=\linewidth]{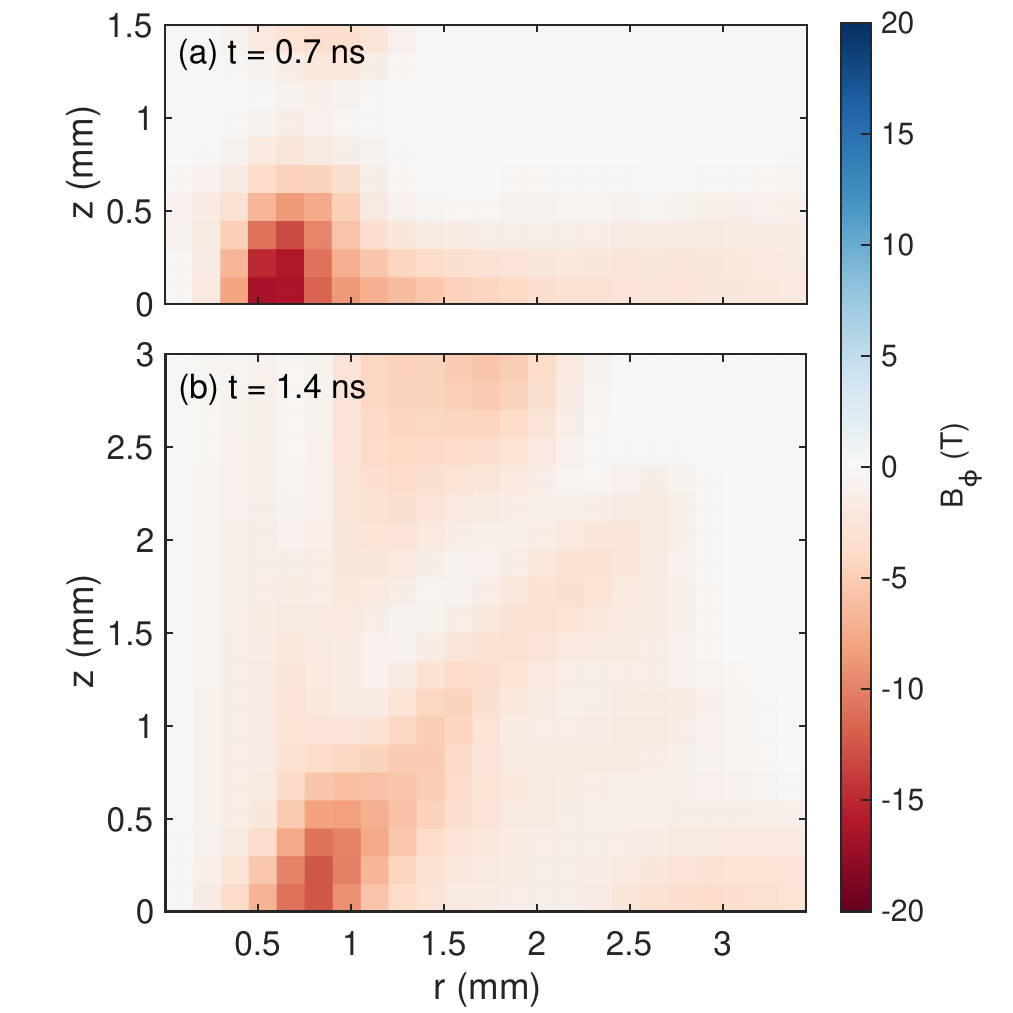}
	\caption{Toroidal magnetic field from experimental tomography at (a) $t=0.7~$ns and (b) $t=1.4~$ns.}
	\label{fig:B_exp}
\end{figure}

Over 3000 proton beamlets were tracked across all views and both timings in this experiment. The tomographic inversion for both timings is shown in Fig. \ref{fig:B_exp}. The two timings reveal a clear evolution of the field structure.

At early time ($t=0.7~$ns), the experimental fields are predominantly located close to the target surface within in the first 300 $\mu$m. The strongest fields occur near $r\approx0.6~$mm, roughly corresponding to the edge of the laser spot and the Biermann ``ring'' feature in the radiographs. In addition to this near-axis peak, there are extended fields out to large radius of $r\approx2$--3~mm, which correspond to weak but non-zero deflections at large radius visible in the radial lineout in Fig. \ref{fig:data_comb}(d).

In contrast, the later time ($t=1.4~$ns) experimental fields extend significantly off the target surface into the coronal plasma and have a complex field structure. The finite height of the fields is qualitatively consistent with the Biermann ring shifts presented in Sec. \ref{sec:qual_tomo}. We find that the coronal fields are \emph{necessary} to reproduce the observed proton deflections; removing coronal fields in the inversion prevents a good match to experiment\cite{griff-mcmahon_structure_2025}. These inversions show that the field structure is strongly time dependent, transitioning from a surface-localized structure at $t=0.7~$ns to a volume-filling coronal structure by $t=1.4~$ns.

Several observations can be drawn from the field evolution in Fig. \ref{fig:B_exp}. 
First, we observe extended coronal magnetic fields at late time that are strong enough to magnetize the plasma (Hall parameter $\Omega_e \tau_e \gg 1$ where $\Omega_e$ is the electron cyclotron frequency and $\tau_e$ is the electron collision time). This differs from multiple MHD simulations that predict the field is located close to the target with minimal coronal extent \cite{lancia_topology_2014,gao_precision_2015,campbell_measuring_2022}. It also differs from the shell-like magnetic-field morphology reported by Li \textit{et al.}\cite{li_measuring_2006}, where the fields were concentrated on a hemispherical shell.

Secondly, the evolution of the field structure gives insights into the field transport. At both timings, the strongest fields occur near the edge of the laser spot and move outwards from $r=0.6$~mm at 0.7~ns to $r=0.8$~mm at 1.4~ns. This corresponds to a speed of 285~km/s, comparable to the plasma sound speed and consistent with advection of the magnetic fields by the bulk plasma flow. Finally, the field distribution expands dramatically off the target surface from early to late time. This behavior is difficult to reconcile with a picture in which Nernst advection continuously anchors the field at the ablation front, and suggests that either magnetic transport into the corona or coronal field generation becomes increasingly important later in the evolution. We stress that the $z$-directed field transport we measure here requires a tomographic analysis and was not visible in prior experiments from a single proton view \cite{li_measuring_2006,petrasso_lorentz_2009,willingale_fast_2010,lancia_topology_2014,gao_precision_2015,campbell_measuring_2022}. 

\subsection{Extended MHD simulations} 

\begin{figure}
	\includegraphics[width=\linewidth]{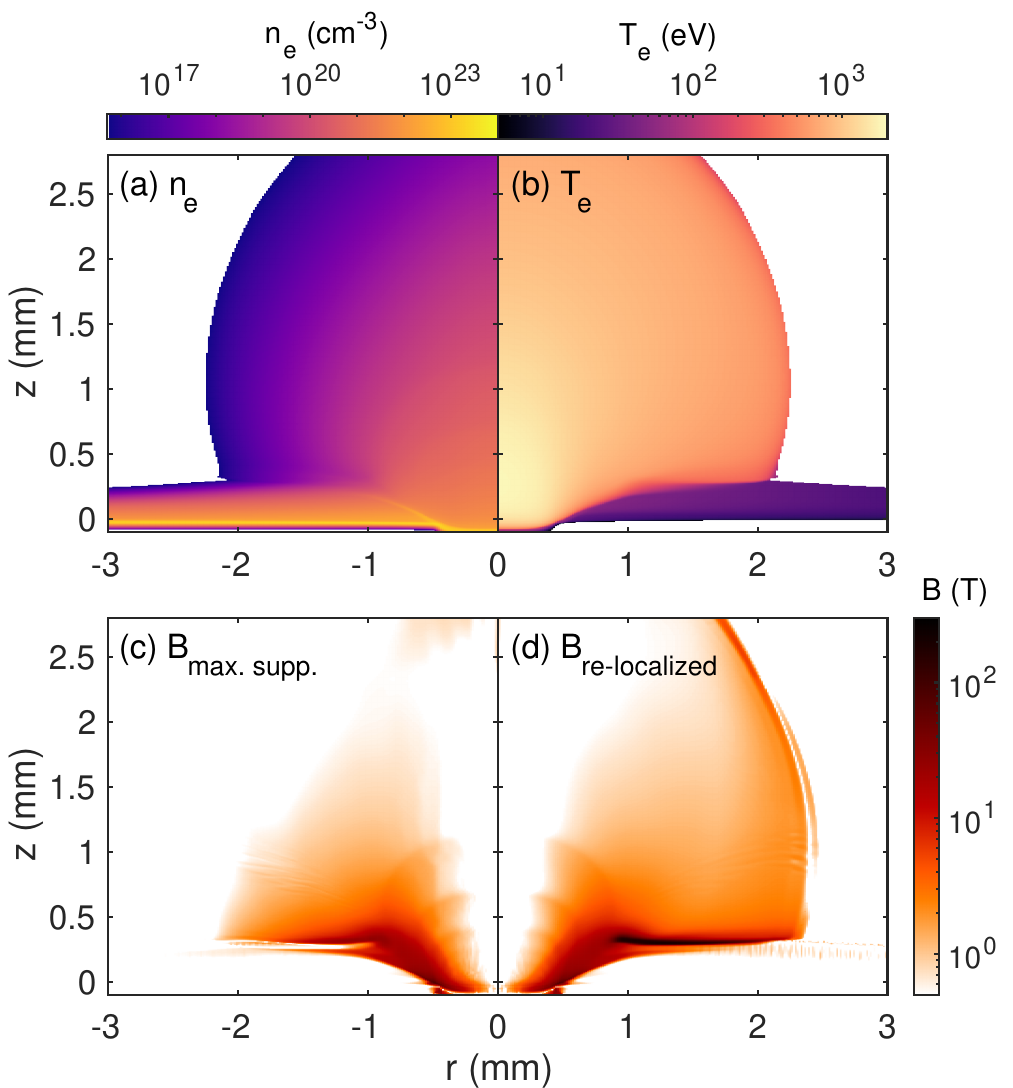}
	\caption{Simulation outputs at $t=1.4~$ns. (a) Electron density. (b) Electron temperature. (c,d) Magnetic field comparing the maximally suppressive Biermann model with the re-localized model, as described in Davies \cite{davies_nonlocal_2023}.}
	\label{fig:GORGON_2x2}
\end{figure}

The experimental fields are directly compared to extended MHD simulations using the \textsc{gorgon} code \cite{ciardi_evolution_2007,chittenden_recent_2009,walsh_extended-magnetohydrodynamics_2020}. The magnetic field in the simulations is described with the extended MHD framework  \cite{epperlein_plasma_1986,braginskii_transport_1965}.
\begin{align}
\frac{\partial \vec{B}}{\partial t}
&= \nabla \times (\vec{v}_B \times \vec{B})
 - \nabla \times \frac{\eta}{\mu_0}\nabla \times \vec{B} \nonumber\\
&\quad
 + \nabla \times f_B \frac{\nabla P_e}{e n_e}
 + \nabla \times \frac{\beta_\parallel \nabla T_e}{e} \label{eqn:xMHD_induction} \\
\vec{v}_B
&= \vec{v}
 - \gamma_\perp \nabla T_e
 - \gamma_\wedge(\hat{\vec{b}} \times \nabla T_e)
 - \frac{\vec{J}}{e n_e}(1+\delta^c_\perp) \nonumber\\
&\quad
 + \frac{\delta_\wedge ^c}{e n_e}(\vec{J} \times \hat{\vec{b}})
\label{eqn:magnetic_advection}
\end{align}
The first term on the right-hand side of Eq.~\eqref{eqn:xMHD_induction} describes effective magnetic-field advection, including bulk fluid advection, electron heat-flux advection (the Nernst effect), and current-dependent contributions given in Eq.~\eqref{eqn:magnetic_advection}. The coefficients $\eta,\,\beta,\,\gamma,$ and $\delta$ are transport coefficients that have recently been updated to more accurately capture transport at low magnetization \cite{davies_transport_2021,sadler_symmetric_2021,walsh_updated_2021}. The Nernst advection term was flux-limited to approximate kinetic suppression of magnetic transport\cite{walsh_kinetic_2024}. The second term in Eq.~\eqref{eqn:xMHD_induction} describes resistive diffusion. 
The last two terms describe field generation from the Biermann-battery effect and from ionization gradients \cite{sadler_magnetic_2020}, although ionization gradients do not contribute significantly in these simulations. The Biermann term includes a suppression factor $f_B$ that accounts for non-local transport when $\lambda_{ei}/L_T \gtrsim0.01$, described later in this section. Integrating Eq.~\eqref{eqn:xMHD_induction} over a surface and applying Stokes' theorem gives the corresponding evolution equation for the enclosed magnetic flux. The $\vec{v}_B \times \vec{B}$ term is conservative and would only lead to change of magnetic flux with finite outward advection at the edge of the domain.

Figure \ref{fig:GORGON_2x2} shows simulation outputs at $t=1.4~$ns of electron density, electron temperature, and magnetic fields using two different models for magnetic suppression. The ``maximally suppressive" model depends solely on the non-locality and is defined in Sec. IV of Davies \cite{davies_nonlocal_2023}. In contrast, the re-localized model in Fig. \ref{fig:GORGON_2x2}(d) accounts for sufficiently strong magnetic fields that localize the transport and therefore counter the suppression, defined in Sec. V of Davies \cite{davies_nonlocal_2023}. In this case, the suppressive factor $f_B$ depends on both the non-locality and the magnetization. 
Physically, this re-localization occurs one the electrons become magnetized ($\Omega_e\tau_e\gg1$). The relevant transverse transport length scale is no longer the collisional mean free path $\lambda_{ei}$, but instead is limited to the electron Larmor radius $\rho_e$. In this regime, the nonlocal electron transport that suppresses the Biermann term is substantially reduced, allowing the Biermann source to approach its classical value.
Interestingly, the re-localization in this simulation is strong enough that the Biermann source is effectively unsuppressed ($f_B\approx1$), yielding the same field generation as a simulation without any Biermann suppression at all. This can be interpreted either as minimal Biermann suppression due to local transport, or as suppression that is mitigated once the self-generated fields magnetize the plasma and re-localize the electron transport.

\begin{figure}
	\includegraphics[width=\linewidth]{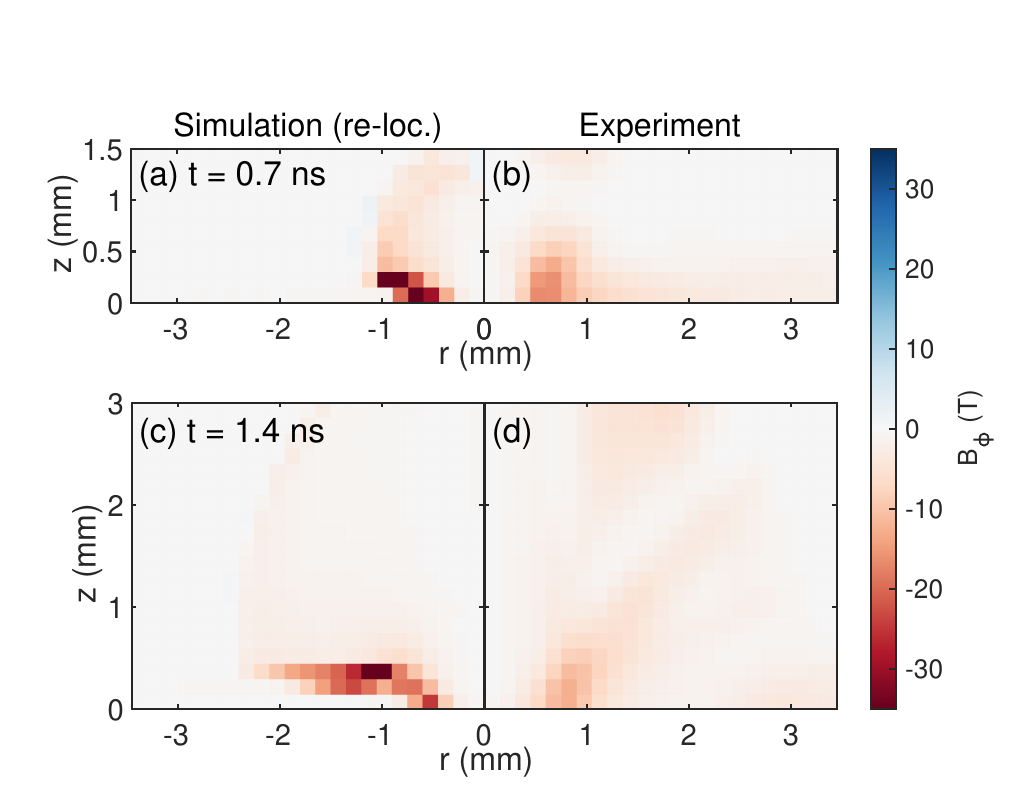}
	\caption{Toroidal magnetic field from extended MHD simulation with re-localized Biermann suppression model (left) and experimental tomography (right). The top row is at $t=0.7~$ns and the bottom row is at $t=1.4~$ns. The simulation was run at high resolution ($\Delta z=7~\mu$m and $\Delta r=14~\mu$m), but was binned and downsized to match the experimental resolution of $150~\mu$m. The target is located at $z=0$ in experiment and simulation.}
	\label{fig:B_evolution}
\end{figure}

The magnetic field in these simulations is mainly generated and located in the ablation front close to the target, where the electron temperature sharply transitions from the hot plume to the cold, dense target. This strong temperature gradient, combined with a non-parallel density gradient, produces the Biermann source term. The fields are then advected and compressed by the Nernst effect near the plasma-target interface. The re-localized simulation has more extended coronal fields and stronger fields near the target interface than the maximally suppressive case.

\begin{figure}
	\includegraphics[width=\linewidth]{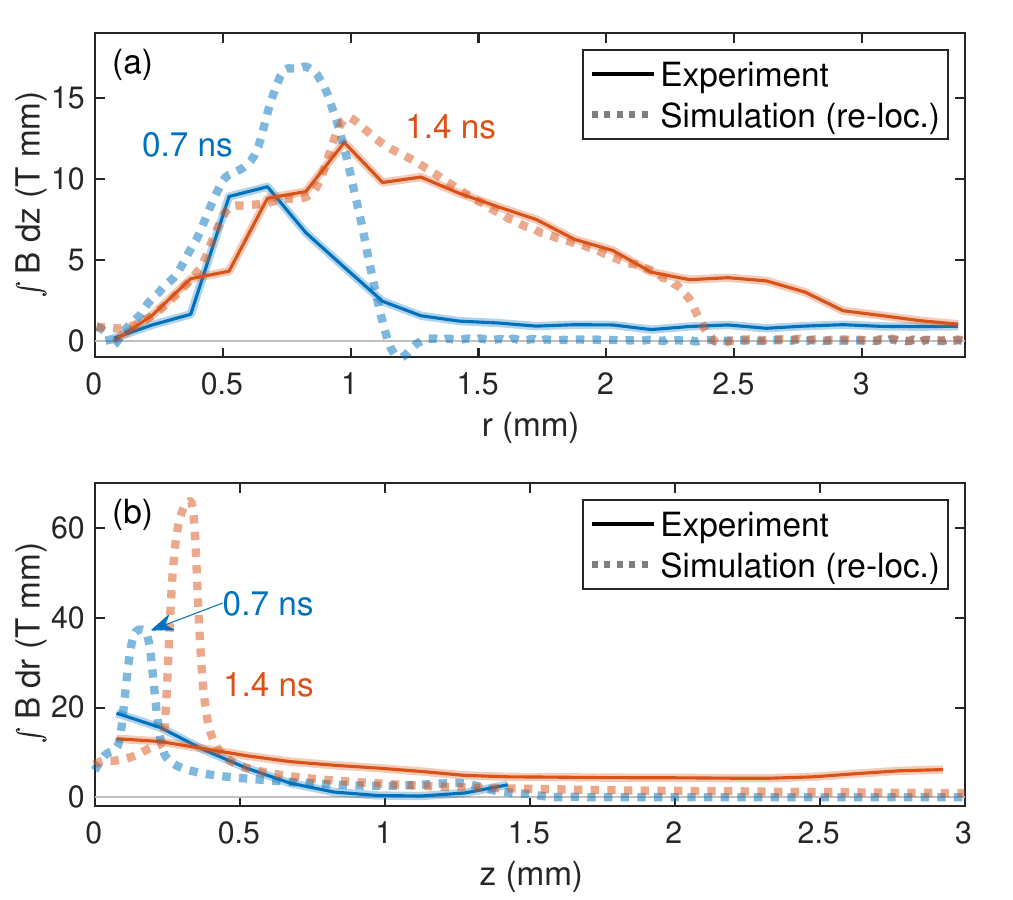}
	\caption{Evolution of the path-integrated magnetic fields in experiment and simulation. (a) Radial profile of $\int B_\phi\,dz$. (b) $z$-profile of $\int B_\phi\,dr$. Blue curves show $t=0.7~$ns and red curves show $t=1.4~$ns. Solid lines refer to experiment and dashed line refer to simulation with re-localized Biermann suppression model.}
	\label{fig:Bdl_exp_vs_sim}
\end{figure}

Figure \ref{fig:B_evolution} compares the experimental fields to the re-localized simulation, block-averaged over the inversion cell size of 150$~\mu$m. At $t=0.7~$ns, the simulation and experiment show similar structure; both contain a strong field feature at $r=0.6~$mm close to the target surface. However, the simulation does not contain the weak fields inferred experimentally at large radius ($r=2.5$--3.5~mm). By $t=1.4~$ns, the magnetic structure diverges more substantially between the simulation and experiment. Most of the magnetic field in the simulation is confined to a thin ``wing" structure at the ablation front at $z=0.4~$mm, with weak coronal fields at the 1 to 2 T level. In contrast, the experimental inversion peaks close to the target and contains appreciable field throughout the corona.

This comparison is made clearer in Fig. \ref{fig:Bdl_exp_vs_sim} by plotting the magnetic fields integrated in the $r$ and $z$ directions  The radial profile, $\int B_\phi\,dz$, shows clear radial expansion of the experimental and simulation fields as the peak field shifts outwards in time. At $t=0.7~$ns, the agreement is poorer because the experimental fields are more radially extended experimental and remain stronger near the edge of the plasma bubble. At $t=1.4~$ns, there is relatively good agreement in the radial profile, although the experimental fields extend to larger radius. The $z$-profile shows more striking differences between experiment and simulation. The simulation fields are consistently peaked near the target whereas the experimental fields originate near the target at $t=0.7~$ns but extend off the target with relative uniformity at $t=1.4~$ns.

\subsection{Magnetic flux evolution}

\begin{figure} [b]
	\includegraphics[width=\linewidth]{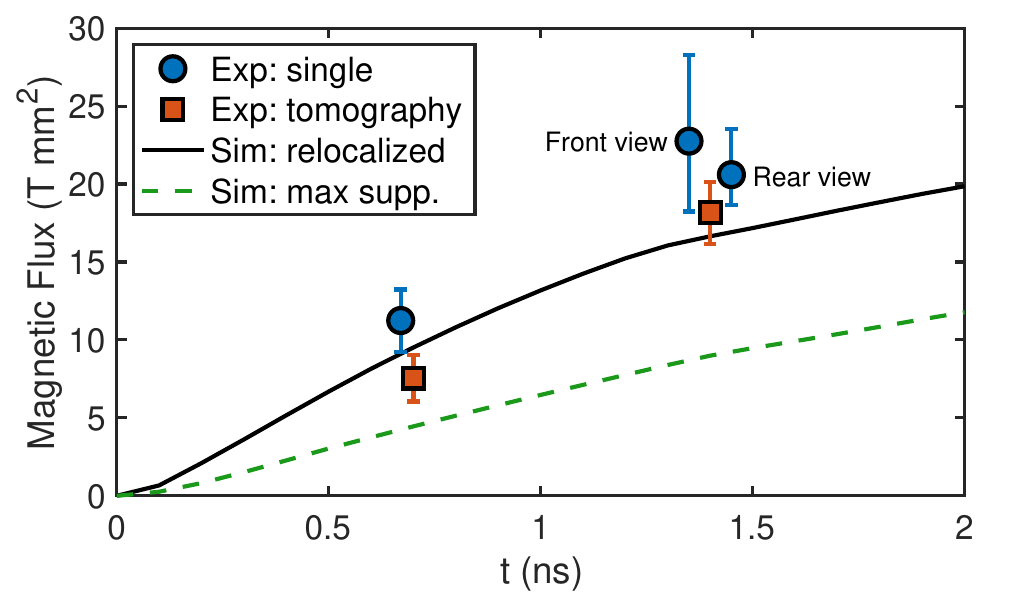}
	\caption{Evolution of the magnetic flux from experiment and simulation. The experimental flux is calculated from single views (front or rear view only; blue circles) or the tomographic inversion (red squares). The simulation flux use either a Biermann suppression model with relocalization (black line) or maximal suppression (green dashed line).}
	\label{fig:BFlux_comp}
\end{figure}

The mesh radiography and tomographic inversion enable direct measurements of the magnetic flux
\begin{equation}
    \Psi(t)=\iint dr\,dz\,B_\phi (r,z,t)
\end{equation}
and its evolution in time. The magnetic flux evolution follows from integrating the induction equation, Eq.~\eqref{eqn:xMHD_induction}, over the measurement area, and includes the effects of conservative transport, Biermann-battery generation, and resistive dissipation. Assuming that there is no flux lost through the boundary and that resistive dissipation is small, the change in flux results simply from the Biermann generation term and therefore provides a valuable constraint on the Biermann-battery physics. Figure~\ref{fig:BFlux_comp} compares the experimental flux evolution and two extended-MHD simulations that use Biermann-battery suppression models containing either re-localization or maximal suppression \cite{davies_nonlocal_2023}. The maximal suppression model reduces the magnetic flux by a factor of nearly 2 in these conditions. The experimental measurements are obtained either from a single view or from the full tomographic inversion. The single view approach uses the front or rear views in the paraxial approximation, treating the measured path-integrated field as an effective $\int B_\phi\,dz$ at each radius and then integrating radially. In contrast, the tomographic measurement evaluates the full 2D integral over the reconstructed $B_\phi(r,z)$.

The error bars in Fig.~\ref{fig:BFlux_comp} were calculated as follows. In the single view, these reflect the non-axisymmetry of the underlying data, and were calculated by binning the path-integrated field in radius, and integrating the azimuthal standard deviation in each bin, radially. We also note that the $t=1.4~$ns values are lower bounds because the fields extend beyond the radiography field of view, reflected by the asymmetric error bars. In the tomographic measurements, the flux uncertainty is estimated from a view exclusion analysis and bootstrap analysis, presented in the next section. 
There is also a systematic trend that single-view measurements of flux tend to exceed the tomographic values. Differences between the two approaches may be expected because the single-view method uses the paraxial approximation, which produces errors for the $\pm20\degree$ field of view in this experiment.

Table \ref{tab:flux_values} summarizes the flux values used in Fig. \ref{fig:BFlux_comp}. The experimental magnetic flux generation rate is roughly $\partial \Psi/\partial t \approx14~$kV. This contrasts with the lower flux growth rate of $\partial \Psi/\partial t \approx4~$kV reported in a prior experiment under similar conditions \cite{campbell_measuring_2022}. The experimental flux most closely matches the re-localized simulation at both timings, supporting minimal suppression of Biermann-battery field generation under these conditions. 
\begin{table}
\caption{\label{tab:flux_values}Magnetic flux $\Psi=\iint B_\phi(r,z)\,dr\,dz$ (T mm$^2$) from experiment and simulation. The $t=1.4~$ns single-view entry lists the front and rear values, respectively.}

\begin{ruledtabular}
\begin{tabular}{l|cc|cc}
Time & \multicolumn{2}{c|}{Experiment} & \multicolumn{2}{c}{Simulation} \\
 & Single & Tomo. & Re-localized & Suppressed \\
\hline
$0.7~$ns  & 11.2 & 7.6  & 9.5  & 4.5 \\
$1.4~$ns  & 22.7, 20.6 & 18.1 & 16.6 & 9.0 \\
\end{tabular}
\end{ruledtabular}
\end{table}

Finally, the increase in coronal magnetic field between the two timings implies either significant magnetic transport away from the target or magnetic field generation in the corona itself. The coronal flux is defined as
\begin{equation}
\Psi_{\mathrm{cor}}(t)= \iint_{z>z_0} dr\,dz\,B_\phi(r,z,t),
\end{equation}
where we have chosen $z_0=0.5~$mm to separate the near-target region from the corona. In the tomographic inversions, the coronal flux increases by $\Delta\Psi_{\mathrm{cor}}\approx 10.8~\mathrm{T\,mm^2}$ between $0.7$ and $1.4~$ns.

The measured change in coronal flux is compared to a simple estimate based on frozen-in fluid advection through the plane at $z=z_0$.
\begin{equation}
\Delta \Psi_{\mathrm{cor}} \approx \int_{0.7\text{ ns}}^{1.4\text{ ns}} dt \int dr\, \left[ V_z(r,z_0,t)\,B_\phi(r,z_0,t)\right]
\label{eq:coronal_adv_est}
\end{equation}
We take the $z$-directed bulk flow $V_z(z_0)\approx1000$~km/s from simulation and the radial integral $\int dr\,B_\phi(r,z_0,t)\approx8.5~\mathrm{T\,mm}$ from both inversions. These values give $\Delta\Psi_{\mathrm{cor}}\approx 6~\mathrm{T\,mm^2}$ from fluid advection alone. The measured change in coronal flux exceeds this value by a factor of ${\sim}2$, suggesting that additional transport or generation mechanisms contribute to the coronal fields at late times.

\subsection{Inversion validation}

\begin{figure} [b]
	\includegraphics[width=\linewidth]{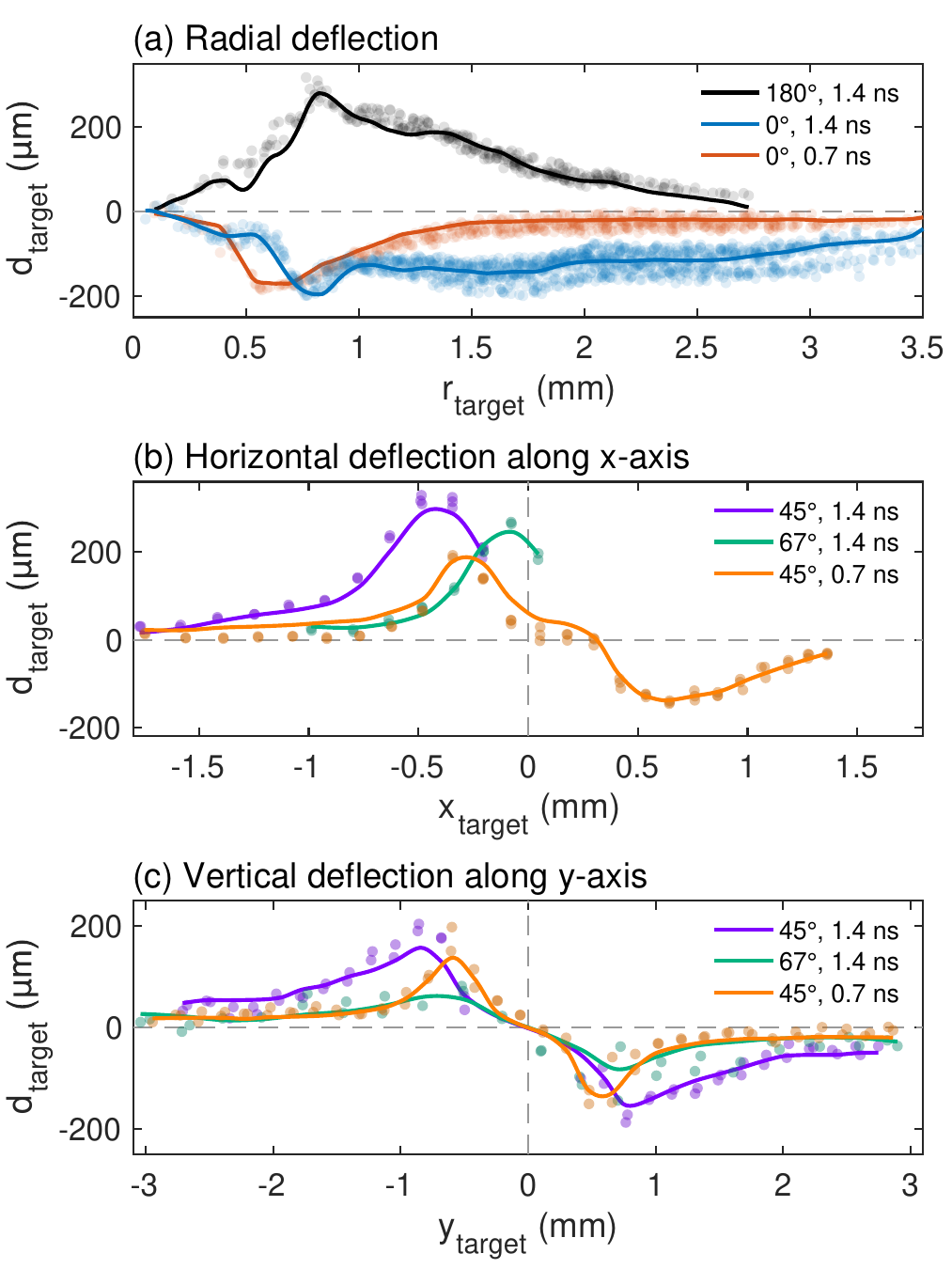}
	\caption{Deflection lineouts of the data (dots) and synthetic deflections from inversions (lines). (a) Radial deflection vs. radius for front and back views. (b) Horizontal deflection along the $x$-axis for oblique views. (c) Vertical deflection along the $y$-axis for oblique views. The deflections and position coordinates have been de-magnified from the detector.}
	\label{fig:def_validation}
\end{figure}

The inversions are validated by forward-modeling the proton deflections from the reconstructed fields and directly comparing to the measured beamlet deflections, summarized in Fig.~\ref{fig:def_validation}. For each view, we compute synthetic deflection maps by propagating protons through the reconstructed $B_\phi(r,z)$, accounting for the full cone-geometry, non-paraxial projection factors, and deflection that occurs before the target plane. Deflection lineouts are extracted along the radial direction for the front and rear views, and along the horizontal and vertical directions for the oblique views. The deflections are de-magnified from the detector to the target units. In all cases, excellent agreement is observed between the measured and synthetic deflection data and leads to high confidence in the final inversion. The normalized root mean square (RMS) error across all deflections is 34.3\% and 24.8\% for $0.7$ and $1.4$~ns, respectively. Much of this error is dominated by either azimuthal variation that is not captured in an axisymmetric system, or azimuthal deflection in the front and rear views that is not captured by a purely toroidal magnetic field.

The vertical lineout in Fig.~\ref{fig:def_validation}(c) also provides a symmetry check. Under the axisymmetric assumptions used in the inversion, the vertical deflection should be antisymmetric across $y=0$. We observe clear antisymmetry that supports both the axisymmetric model and that the co-registration of the origin across the images was correct.

\begin{figure}
	\includegraphics[width=\linewidth]{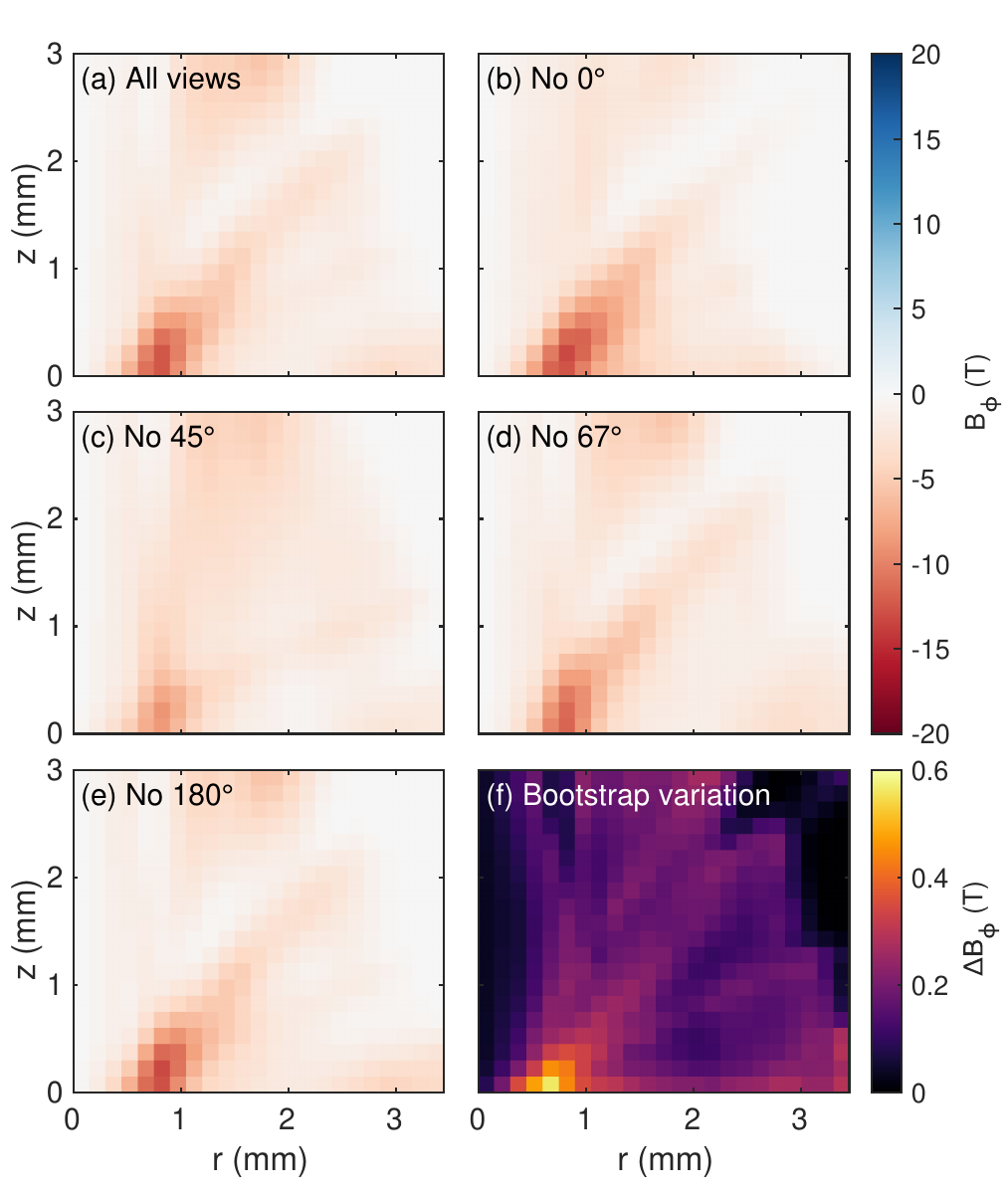}
	\caption{Inverted magnetic fields from (a) including all views and (b-e) from excluding a single view. (f) Standard deviation in $10^4$ inversions from bootstrap sampling with replacement, using all views.}
	\label{fig:B_unc}
\end{figure}

To test sensitivity to the finite number of view angles, we repeated the inversion four times at $t=1.4$~ns, excluding a single view each time, which we call the ``exclusion analysis". These are shown in Figs.~\ref{fig:B_unc}(a-d) and contain similar field structure; the magnetic field extends significantly off the target into the corona in all inversions. However, the striations in the data vary more significantly across the inversions, suggesting that these small-scale features have higher uncertainty and are less robust.

We also tested sensitivity to individual deflection measurements by bootstrap sampling \cite{efron1994introduction}. This technique samples the deflection data with replacement to produce many new datasets of the same size as the original. $10^4$ inversions were produced using bootstrapped samples and their standard deviation is plotted in Fig. ~\ref{fig:B_unc}(e). The maximum standard deviation across the inversions was $0.6$~T and was located at the peak field region (close to the target at $r\approx0.75~$mm), corresponding to a 5\% variation in field strength. The coronal fields remained comparatively stable and were insensitive to removing individual deflection measurements. 

\begin{figure}
	\includegraphics[width=\linewidth]{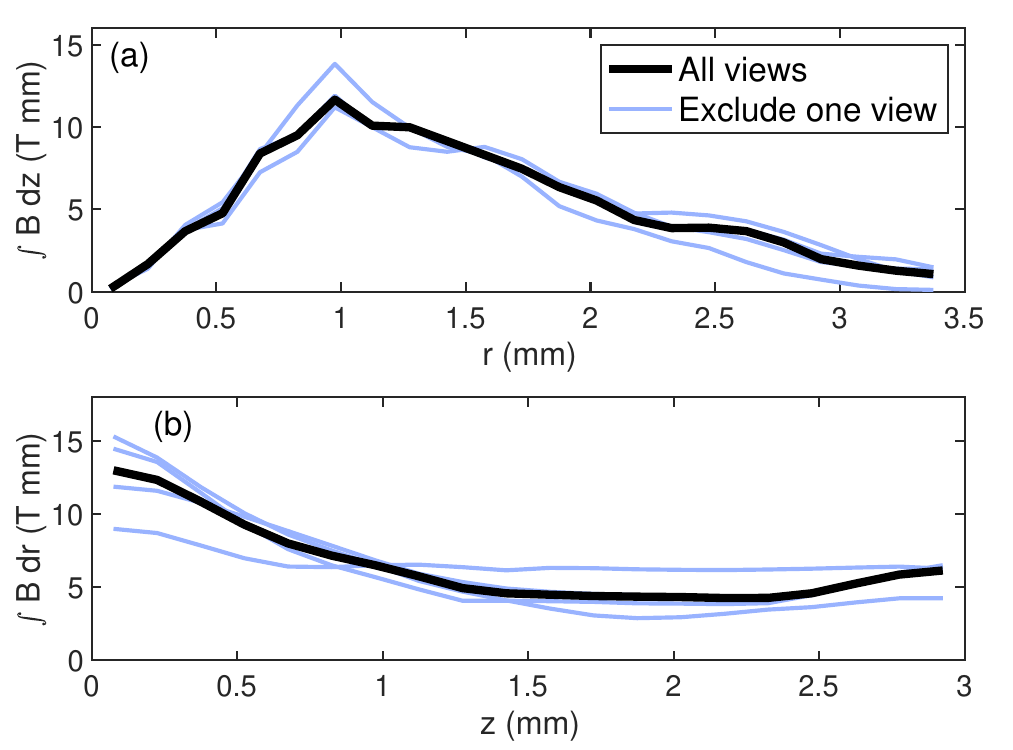}
	\caption{Path-integrated experimental magnetic fields at ${t=1.4~}$ns. (a) Radial profile of $\int B\,dz$. (b) $z$-profile of $\int B\,dr$. The black line includes all views while the light blue lines exclude a single view in the inversion.}
	\label{fig:Bdl_exp_unc}
\end{figure}

The robustness across the exclusion analysis is further visualized by plotting integrated profiles in Fig.~\ref{fig:Bdl_exp_unc}. The profiles compare inversions that exclude different views (thin blue lines) and an inversion that includes all views (thick black line). There is good agreement in both the radial and $z$-profiles across the inversions. The maximum variation is 2~T$\,$mm in the radial profile (17\% of the maximum value) and 4~T$\,$mm in the $z$-profile (30\% of the maximum value). The field at large $z$ remains non-negligible, indicating that extended coronal magnetization is a robust feature rather than an artifact of a particular view or a small subset of deflection measurements. Most importantly, the differences relative to the MHD simulation (Fig.~\ref{fig:Bdl_exp_vs_sim}) are robust with respect to the view exclusion analysis.

\section{Discussion and conclusion}

We report measurements of the magnetic field structure and magnetic flux at two different timings in a laser-solid interaction using proton tomography and high-fidelity mesh radiography. We find that the experimental fields evolve from being localized near the target at early time ($t=0.7~$ns) to extending off the target into the corona at late time ($t=1.4~$ns), where they are strong enough to magnetize the plasma. The Hall parameter magnetization scales as $\Omega_e \tau_e \sim B T_e^{3/2} n_e^{-1}$ so that even relatively weak fields can magnetize the hot and rarefied corona. The Hall parameter controls several key transport processes, including suppression of the electron flux perpendicular to the field $[\kappa_\perp / \kappa_\parallel \approx 1/(1+\Omega_e^2 \tau_e^2)]$ and deflection of the heat flux by the Righi-Leduc effect $[\kappa_\wedge / \kappa_\parallel \approx \Omega_e\tau_e/(1+\Omega_e^2 \tau_e^2)]$ \cite{braginskii_transport_1965}. The experimental volume-averaged field in the corona is 1.8~T$\,$mm$^2$, which corresponds to $\Omega_e \tau_e=9$ for typical coronal parameters of $n_e=10^{19}~$cm$^{-3}$ and $T_e=1~$keV.  The magnetization continues to increase further into the corona as the density further decreases, but the field and temperature remain relatively constant. This magnetization level is already sufficient to reduce the heat flux perpendicular to the magnetic field by ${\sim}$97\% compared to the unmagnetized case \cite{sadler_symmetric_2021}. Magnetization insulates the plasma and leads to stronger temperature gradients in the corona, directly impacting the global plasma behavior and processes like laser-plasma coupling \cite{farmer_simulation_2017,lezhnin_simulations_2025}.

The experimental results are compared to extended MHD simulations. At early timing, the experiment and simulation show good agreement in the field structure, reproducing both the peak location near the edge of the laser spot and the overall field morphology. At later time, the field structure has only moderate agreement; the experiment and simulation both contain coronal fields at the 1 to 2 T level, but the structure near the target differs significantly.  In the simulation, the field remains compressed into a thin layer at the ablation front whereas the experimental field is much weaker and extends throughout the domain. The experimental coronal fields are robust to a range of sensitivity tests and indicates that these fields are physical rather than an artifact of the inversion. 

A comparison of the magnetic flux shows good agreement with the simulation that includes the re-localized model for Biermann-battery suppression and poor agreement with the maximally suppressive model. A key point is that the magnetic flux isolates net field generation whereas the field structure is also influenced by conservative advection of the fields. This suggests that the re-localized Biermann model describes the field generation well, but the magnetic transport models still need improvement.

One plausible interpretation is that the simulation over-anchors the field to the ablation front through overly strong Nernst advection.
An intuitive picture is that the field distribution in the target-normal direction is set by competition between outward bulk-flow advection and inward Nernst advection, producing a transition layer where $|v_{\rm plasma}|\simeq|v_{\rm Nernst}|$ discussed in Refs.~ \cite{nishiguchi_nernst_1985,manuel_instability-driven_2013}. Fields generated outside this layer are swept outward into the corona by the expanding flow, whereas fields generated inside are drawn back toward the ablation front and target. Reducing $v_{\rm Nernst}$ shifts this balance and naturally increases the coronal field extent. In these simulations, the Nernst advection was flux-limited \cite{walsh_kinetic_2024} but a more complete inclusion of kinetic effects may be necessary to increase the coronal field extent without affecting magnetic flux. Future tomographic measurements at different timings and with varied experimental conditions would help to constrain the magnetic transport. 

More broadly, proton tomography is a highly unexplored diagnostic area in HED physics and enables direct access to the 3D structure of electromagnetic fields. These results illustrate that proton tomography can provide quantitative benchmarks for magnetized transport in HED plasmas. However, most magnetized HED systems that have only been studied with a single radiography line of sight would benefit from a tomographic approach, including magnetic reconnection\cite{nilson_bidirectional_2008,rosenberg_slowing_2015,valenzuela-villaseca_x-ray_2024} and self-generated magnetic fields in hohlraums\cite{li_observations_2009,pearcy_revealing_2025}.

A ripe area for future work is proton tomography based on fluence inversions\cite{schaeffer_proton_2023,bott2017proton}, rather than the mesh deflection method used in this work. Fluence-based inversions provide proton deflections at high spatial resolution based on the proton fluence patterns and capture small scale structures such as the magnetic spokes observed around the laser spot in laser-solid interactions \cite{gao_precision_2015,campbell_magnetic_2020,sutcliffe_observation_2022,griff-mcmahon_proton_2024}. When applied to tomography, fluence inversions would generate more proton deflection measurements and would therefore have better coverage and constraints, but at the cost of reduced accuracy in the deflections themselves \cite{griff-mcmahon_proton_2024}. 
    
The use of fluence inversions also motivates non-axisymmetric inversions for full-3D field structure. The assumption of axisymmetry in this tomographic inversion was necessary due to the reduced number of views and measured deflections, but the tomographic framework could be applied to non-axisymmetric fields in general. Full-3D inversions require more measurements to ensure sufficient coverage. This likely requires more view angles, which could be reduced by using proton fluence inversions as compared to mesh measurements.

It is important to note that non-axisymmetric, fluence-based tomographic inversions are more susceptible to shot-to-shot variations, since fine-scale structure is often seeded by stochastic processes. A direct tomographic reconstruction is only meaningful when the dominant structures are repeatable from shot to shot. If the field evolution is stochastic, a tomographic analysis can still constrain the location of the structures and statistical properties like the RMS fluctuation level, the structural orientation, and dominant wavevector.

In conclusion, we use multi-view proton radiography to reconstruct the time evolution of self-generated magnetic fields in a laser-foil interaction. The tomographic reconstructions show a clear transition from fields anchored to the target at $t=0.7~$ns to substantially more extended coronal fields at $t=1.4~$ns, supported by qualitative arguments. These fields are strong enough to magnetize the coronal plasma which would significantly suppress heat flux. The measured magnetic flux evolution shows best agreement with extended MHD simulations that use a re-localized Biermann-battery suppression model, equivalent to eliminating suppression entirely. However, comparisons of the field structure show only moderate agreement. This combination of results suggests that the field generation model is consistent with experiment, while field transport models require additional development to reproduce the experimental field structure. 
These results are important for ICF hohlraums because self-generated fields in the wall blowoff and near the laser-entrance hole can modify electron heat flow and the underdense plasma conditions that set laser propagation and absorption, which will modify laser-plasma coupling and drive symmetry. A complete understanding of magnetic field generation and transport is also necessary for accurate interpretation of magnetized laboratory astrophysics experiments. Proton tomography opens a new route to directly probe 3D magnetic field structure in laser plasmas.

\section*{Acknowledgements}
The authors thank the MIT and LLE teams for processing the CR39 data. The work was performed under the auspices of the U.S. Department of Energy by General Atomics under NNSA Contract 89233124CNA000365 and by Lawrence Livermore National Laboratory under Contract DE-AC52-07NA27344. The experiment was conducted at the Omega Laser Facility with beam time through the National Laser Users' Facility user program. This work was supported by the Department of Energy under grant Nos. DE-NA0004034, DE-NA0004271, and DE-NA0003868 and by the University of Rochester ``National Inertial Confinement Fusion Program" under award No. DE-NA0004144. This work is supported by the National Science Foundation Graduate Research Fellowship Program under grant No. 2039656.

\appendix

\section{Geometric corrections for mesh} \label{sec:geometric_corr}
\begin{figure}
	\includegraphics[width=0.9\linewidth]{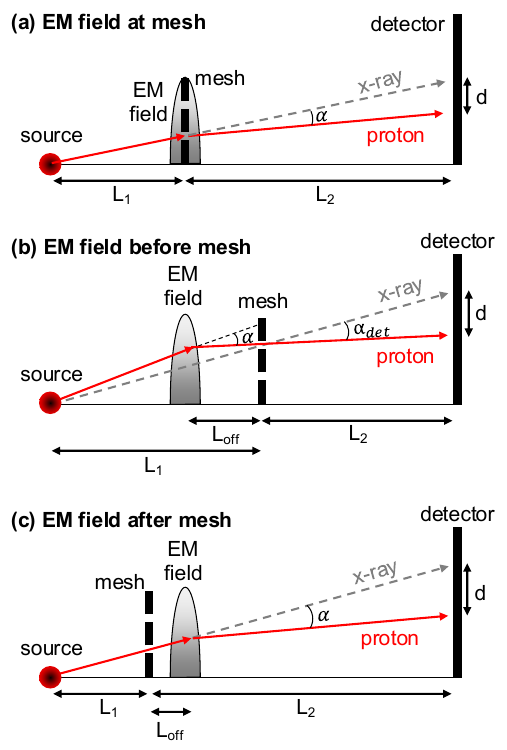}
	\caption{Proton radiography setups with EM fields located (a) at the mesh, (b) before the mesh, and (c) after the mesh. If EM fields are before, a correction of order $L_{\rm{off}}/L_1$ is needed to relate the detector deflection $d$ to the true deflection angle $\alpha$. If EM fields are after the mesh, there is a small correction of order $L_{\rm{off}}/L_2$.}
	\label{fig:MeshSetup}
\end{figure}

The following discussion expands on Ref.~\cite{griff-mcmahon_structure_2025} and is included here for completeness. In mesh proton radiography, the mesh splits the proton beam into individual beamlets that are tracked on the detector. The mapping between the measured beamlet displacement and the underlying electromagnetic (EM) deflection angle depends on where the EM fields are located relative to the mesh (Fig.~\ref{fig:MeshSetup}). We define $L_1$ as the proton source-to-mesh distance and $L_2$ as the mesh-to-detector distance.

We first consider an idealized ``pancaked" geometry in which the EM fields are thin compared with $L_1$ and $L_2$ and are effectively coincident with the mesh plane [Fig.~\ref{fig:MeshSetup}(a)]. The fields deflect the proton with an angle $\alpha$, after wich the proton travels in a straight line to the detector.  The proton deflection on the detector is
\begin{equation}
    d = L_2 \tan\alpha \approx L_2\,\alpha.
    \label{eq:d_alpha_L2}
\end{equation}
where the small-angle approximation is used. In this limit, no geometric correction is required.

The shots discussed in the main text use a scheme where the mesh is attached to the rear surface of the target. This setup efficiently incorporates a mesh into the experiment without requiring an additional OMEGA Ten-Inch Manipulator (TIM). In some configurations, protons encounter the fields before passing through the mesh [Fig.~\ref{fig:MeshSetup}(b)]. This complicates the analysis because protons and x-rays may take different initial trajectories before passing through the same mesh hole. The detector therefore measures an effective exit angle through the mesh $\alpha_{\rm det}=d/L_2$, where $\alpha_{\rm det}\neq \alpha$. This requires a geometric correction that accounts for deflections occurring upstream of the mesh.

We consider a thin field region located a distance $L_{\rm off}$ upstream of the mesh [Fig.~\ref{fig:MeshSetup}(b)]. From similar triangles, the correction is
\begin{equation}
    \alpha_{\rm det} = \alpha\left(1 - \frac{L_{\rm off}}{L_{1}} \right).
    \label{eqn:meshCorr1}
\end{equation}
For example, $L_{\rm off}=2$~mm and $L_1=10$~mm gives $\alpha_{\rm det}=0.8\,\alpha$ which is a 20\% underestimate if the correction is neglected.

If there are multiple deflections at different distances before the mesh, Eq. \eqref{eqn:meshCorr1} extends to
\begin{equation}
    \alpha_{\rm det} = \sum_i\alpha_{i}\left(1 - \frac{L_{\rm off,i}}{L_{1}} \right) \label{eqn:meshCorrAll}
\end{equation}
where $\alpha_i$ is the $i^{th}$ field deflection at a distance $L_{\rm off,i}$ before the mesh.  Eq. \eqref{eqn:meshCorrAll} is implemented in the tomography scheme to account for geometries where the proton passes through the field region first, and then through the mesh.

Finally, there is also a small correction if the EM field is downstream of the mesh [Fig.~\ref{fig:MeshSetup}(c)]. If the field is located a distance $L_{\rm off}$ after the mesh, then the remaining distance to the detector is reduced and
\begin{equation}
    d = (L_2-L_{\rm off})\,\alpha = L_2 \,\alpha \left(1-\frac{L_{\rm off}}{L_2}\right).
    \label{eq:d_alpha_L2}
\end{equation}
This correction is of order $L_{\rm off}/L_2$ and is much smaller than the upstream correction for the typical parameters used in proton radiography ($L_2 \gg L_1$). This correction has not been included in the inversion because it is at the 1--2\% level for our geometry.

\section{Validation with synthetic data} \label{sec:Tomo_Validation}

\begin{figure}
	\includegraphics[width=\linewidth]{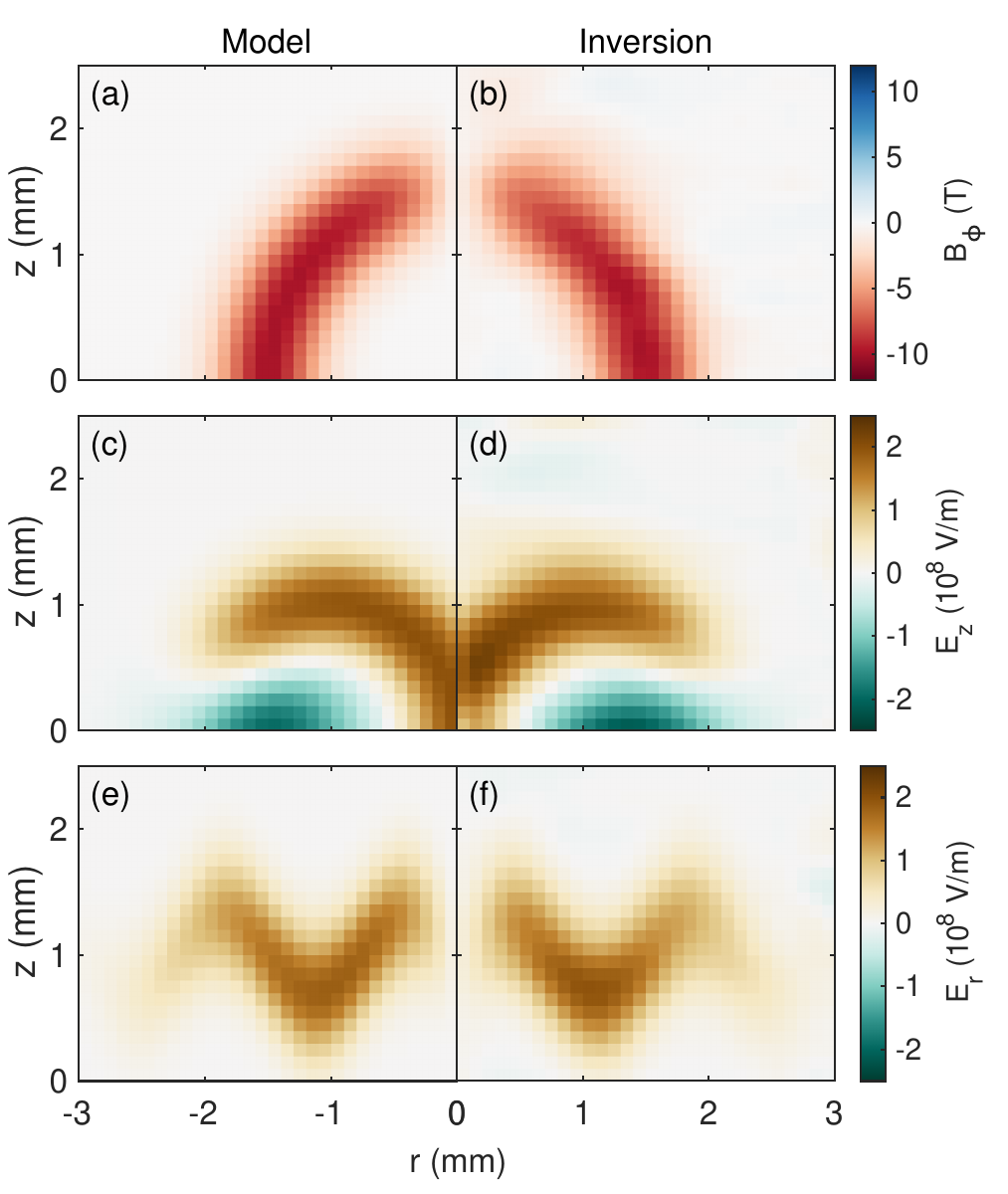}
	\caption{Comparison of model fields (left) and inverted fields using synthetic deflection maps from the four experimental view angles (right). All three field components ($B_\phi$, $E_z$, and $E_r$) were included to generate the deflection maps and were decoupled in the inversion.}
	\label{fig:TomoValidation}
\end{figure}

The following validation is adapted from Ref. \cite{griff-mcmahon_structure_2025} and is included here for completeness. The tomographic inversion procedure was validated using synthetic proton deflection maps from each of the four experimental view angles ($0\degree$, $45\degree$, $67\degree$, and $180\degree$). Three axisymmetric fields, $B_\phi$, $E_z$, and $E_r$, were included to generate the deflection maps, with profiles shown in the left column of Fig. \ref{fig:TomoValidation}. The magnetic profile is a hemispherical shell centered at the origin, the z-directed electric field is part of a toroidal shell with a central feature of the opposite sign, and the radial electric field is an undulating line of field located at a height of $z=1$ mm above the target. These field strengths and profiles were chosen to test possible failure modes of the experimental inversion and investigate the effects of fields with large radius, large height, sign reversals, and overlap between the different fields.

The right column of Fig. \ref{fig:TomoValidation} shows the inverted fields. Excellent agreement is observed in all cases with clear distinctions between regions with and without imposed fields. The inversion was able to isolate the electric and magnetic fields despite significant spatial overlap. The root mean squared error between the model and inversion fields is 0.37 T, $1.1\times10^7$ V/m, and $6.2\times10^6$ V/m for $B_\phi$, $E_z$, and $E_r$, respectively. These correspond to relative errors of 3.7\%, 5.7\%, and 3.1\%, compared to the characteristic field strengths in each profile. These small errors are further reduced when weaker fields are used; protons are assumed to sample the fields along straight-line trajectories, which breaks down when the fields are sufficiently strong and extended along the proton path, discussed in Section \ref{sec:inv_assumptions} in the main text.

\bibliography{Prad_Tomo}

\end{document}